\documentclass[prb,aps,11pt]{revtex4-1}

\pdfoutput=1

\usepackage{amsmath}
\usepackage{amsfonts}
\usepackage{amssymb}
\usepackage{graphicx}
\usepackage{color}
\usepackage{bm}

\usepackage{chngcntr}
\usepackage{mathrsfs}
\usepackage{appendix}

\newcommand*{\citen}{}
\DeclareRobustCommand*{\citen}[1]{%
  \begingroup
    \romannumeral-`\x 
    \setcitestyle{numbers}%
    \cite{#1}%
  \endgroup
}

\newcommand{\changescolor}{black}

\begin{document}

\title{Fermi liquid theory and divergences of the two-particle irreducible vertex in the periodic Anderson lattice}
\author{Corey Melnick$^a$ and Gabi Kotliar$^{a,b*}$}
\affiliation{\textsf{(a)} Condensed Matter Physics and Materials Science Department, Brookhaven National
Laboratory, Upton, NY 11973, USA}
\affiliation{\textsf{(b)} Department of Physics and Astronomy, Rutgers University, New Jersey 08854, USA}
\affiliation{*kotliar@physics.rutgers.edy}

\begin{abstract}
\textbf{Abstract}: Here we analyze the divergences of the irreducible vertex function in dynamical mean field theory, which may indicate either a non-physical breakdown of the perturbation theory or a response to some physical phenomenon. To investigate this question, we construct a quasiparticle vertex from the diverging irreducible vertex functions. This vertex describes the scattering between quasiparticles and quasiholes in a Fermi liquid. We show that the quasparticle vertex \textit{does not} diverge in the charge channel, wherein the irreducible  vertex \textit{does} diverge; and we show that the quasiparticle vertex \textit{does} diverge in the spin channel, wherein the irreducible vertex \textit{does not} diverge. This divergence occurs at the Mott transition wherein the Fermi liquid theory breaks down. Both Hubbard and Anderson lattices are investigated. In general, our results support that the divergences of the irreducible vertex function do not indicate a non-physical failure of the perturbation theory. Instead, the divergences are the mathematical consequence of inverting a matrix (the local charge susceptibility) which accumulates increasingly negative diagonal elements as the Hubbard interaction suppresses charge fluctuations. Indeed, we find that the first symmetric divergences of the irreducible vertex in both Hubbard and Anderson lattices occurs near the maximum magnitude of the (negative) vertex-connected part of the charge susceptibility, which suppresses charge fluctuations.
\end{abstract}

\maketitle

\section{Introduction}

Landau Fermi Liquid theory\cite{Nozieres} is one of the cornerstones of modern quantum many-body theory.

It states that the low energy excitations of a Fermi system can be thought in terms of quasiparticles and quasiholes, with the same quantum numbers the non-interacting Fermi system, but with renormalized parameters such as their mass and velocity. There are residual interactions among those quasiparticles which have to be taken into account when the response to external fields are considered, and they define the Fermi liquid parameters.

The Landau Fermi Liquid theory was later  derived  by diagrammatic perturbation theory. In this derivation, as  intermediate steps,  one utilizes Green's functions which are assumed to have a pole at the quasiparticle energy, and irreducible  (with respect to pairs of Green's functions)  vertex functions which describe the quasiparticle interactions.\cite{AGD,Nozieres}

Landau Fermi liquid was then derived non-perturbatively, using Wilson’s renormalization group,  showing that its validity does not rely on the convergence of perturbation theory.\cite{Shankar1994,Benfatto1995}  Dynamical Mean Field Theory (DMFT) is a non-perturbative technique which is exact in the limit of infinite dimensions, as introduced by Metzner and Vollhardt.\cite{Metzner1989_dmft} Its application to the metallic phases of the Hubbard model enabled the evaluation of the parameters of the Fermi liquid phase and their evolution as the Mott transition is approached. It was therefore a surprise when an important paper\cite{Schafer2013} reported that one of the essential elements in the diagrammatic derivation of Fermi Liquid theory,  the irreducible vertex in the particle-hole charge channel, diverges in DMFT well before the transition is reached.

A divergence computed within a theoretical approach can either indicate a physical effect or an unphysical breakdown of an approximation.  For example, the susceptibility of the order parameters diverges at a second order phase transition.  In this case the divergence is a physical effect, and one learns something from the divergence in a perturbative calculation of a second order phase transition above its upper critical dimensionality.  In contrast, one finds unphysical divergences in the spin susceptibility of a Kondo impurity due to the breakdown of perturbation theory, which  indicates the need for more advanced theoretical techniques such as the renormalization group  technique.

The divergence of the irreducible vertex was investigated further in Refs. [\citen{Schafer2016,Gunnarson2016,Gunnarson2017,Chalupa2018}]. While it was first interpreted as a precursor to the Mott transition, it was later shown that the divergence also occurs in the Anderson impurity model, which does not undergo a Mott transition. Therefore, the divergence was reinterpreted as a growth of the negative, diagonal components of the local susceptibility matrix due to the suppression of charge fluctuations by the Hubbard interaction\cite{Gunnarson2016}. 
\textcolor{\changescolor}{ Additionally, 
Nourafkan \textit{et al}. recently reported that the Hubbard model can undergo a Mott transition without a divergence occurring in the irreducible vertex, provided  the particle-hole symmetry is broken.\cite{Nourafkan2019} They found that the eigenvalues of the local susceptibility become complex when this symmetry is broken, and the imaginary parts of these eigenvalues prevent the divergences from manifesting as the real part of the eigenvalues cross zero.\cite{Nourafkan2019}
}

With this background established, let us outline the objectives of this study.  First, we want to lay to rest the worries that the divergence of the vertex function is  an artifact of DMFT and establish that it instead reflects the physical nature of charge fluctuations in the strongly correlated regime.  Second, we want to evaluate the Fermi liquid interaction parameters and show that they remain finite despite the divergence of the irreducible vertex, even when the irreducible vertex is used as a fundamental building block in the Fermi liquid theory. Finally, we want to demonstrate the generality of the divergence of the irreducible  charge vertex function in the strong correlation regime. To accomplish these goals, we study the periodic Anderson lattice model (PAM) within DMFT and compare the results with the Anderson impurity model (AIM) and Hubbard model. 

\begin{figure}
\includegraphics[scale=1]{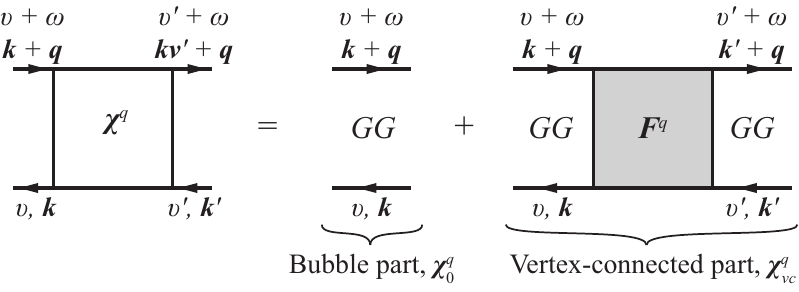}
\caption{\textcolor{\changescolor}{
The total susceptibility, $\chi$, is decomposed into the part generated by bubble diagrams, $\chi_0$, and part generated by the remaining, vertex-connected parts, $\chi_{vc}$. 
}}\label{fig:vertex-connected parts}
\end{figure}

We show that the divergences of the Anderson impurity, as found in Ref. [{\citen{Chalupa2018}}], also occur in the Anderson lattice and correlate with the suppression of charge fluctuations. Indeed, we show that the first symmetric divergence occurs near the maximum  magnitude of the negative vertex-connected part of the charge susceptibility in both the Anderson and Hubbard lattices.  (Figure \ref{fig:vertex-connected parts} shows the decomposition of the susceptibility into bubble and vertex-connected parts.) To understand why these divergences are non perturbative, let us make an analogy between the irreducible vertex and the self energy, which is the one-particle, rather than two-particle, irreducible object.  The divergence of the self energy (via a pole in the DMFT Green's function) signals  the opening  of a gap  or pseudogap in the one particle spectrum.  Similarly, the divergence of the vertex function  reflects the non-perturbative suppression of the charge fluctuations. It is important to recognize that the vertex function, just like the self energy, is only an intermediate object, which in itself is devoid of direct physical interpretation: The evaluation of a self energy is just an intermediate step in the computation of the physical spectral function, and the evaluation of the irreducible vertex is just an intermediate step in the computation of the physical response functions. 

A  local  ($k$-independent) self energy  leads to a description of the transport  via quasiparticles with a modified dispersion ($k$-dependent) relative to the bare particles.  Similarly,  the effect of a local, $k$-independent but strongly $\omega$-dependent vertex function was shown to involve a frequency independent  but $k$-dependent vertex function at low energies. (For recent examples of these effects see Refs.  [{\citen{Wenhu2013}}] and [{\citen{Hyowon2011}}].) This vertex function leads to a description of the interaction of the re-normalized quasiparticles. Here we aim to show that the divergence in the (non-physical) vertex function does not indicate a divergence in the (physical) quasiparticle interaction and the failure of the perturbation theory.  

A valid Fermi liquid theory requires continuity in the coupling constant and adiabatic continuity (lack of level crossings) from the non-interacting case. This is a weaker requirement than analyticity in the coupling constant  and convergence of the perturbation theory series. Still, we support our physical interpretation of the divergence in the perturbation theory by deriving and evaluating the Fermi liquid interaction parameters (and a quasiparticle vertex). In our formalism, the full vertex, $\bm{F}^{q}$, is an expansion of all quasiparticle-quasihole interactions:
\begin{align}
\bm{F}^{q} =  \bm{\Gamma}_{qp}^{q} +  \bm{\Gamma}_{qp}^{q} \bm{\chi}^{q}_{0,qp}\bm{F}^{q},
\end{align}
where $\bm{\Gamma}_{qp}^{q}$ is the quasiparticle vertex and $\bm{\chi}^{q}_{0,qp}$ are the dressed quasipraticle-quasihole lines. As we will show, one can extract the antisymmetric and symmetric, static Fermi liquid parameters, $A_0^{(a/s)}$, from the quasiparticle vertex in the spin and charge channels,
\begin{align}
A_0^{(a/s)} &= z^2 D^*(0)\Gamma_{qp}^{(m/s)}(\bm{q}=0,i\nu_{-1},i\nu_{-1},i\omega_{1}) 
\end{align}
where $z^2$  is the weight of the  quasiparticle, $D^*(0)$ is the reduced density of states at the Fermi level, and $\nu_n$ and $\omega_m$ are Fermionic and Bosonic matsubara frequencies. As the quasiparticle vertex describes a physical interaction in a Fermi-liquid, it and the associated Fermi-liquid parameters should be well behaved in the Fermi-liquid regime.

 We show that the symmetric Fermi liquid parameter and the associated quasiparticle vertex in the charge channel do not diverge, despite the numerous divergences in the associated irreducible vertex. Furthermore, we show that the antisymmetric Fermi liquid parameter and the associated quasiparticle vertex in the spin channel \textit{do} diverge, but only when the Fermi-liquid is no longer well defined, e.g., at a Mott metal-to-insulator (MIT) transition. Note that the irreducible vertex does diverge in the charge channel, but it does not diverge in the spin channel.\cite{Schafer2013,Schafer2016,Gunnarson2016,Chalupa2018} Therefore, we show that our Fermi liquid theory captures the breakdown of the Fermi liquid theory at the MIT, and does not reflect underlying divergences in the irreducible vertex used to build the quasiparticle vertices.  We stress the important role of the incoherent parts of the Green's function in avoiding the divergence and show how Fermi liquid theory provides guidance when interpreting the perturbative results. 
 
 \textcolor{\changescolor}{
Krien \textit{et al.} recently investigated the Fermi liquid parameters in the Hubbard model within DMFT\cite{Krien2019_fermiliquid}. Despite investigating a different lattice and developing a different formalism, our results agree surprisingly well.}

In Sec. \ref{sec:Formalism} we will derive the formalism. Then, in Sec. \ref{sec:PAM results} we will present one-particle results for the PAM, and in Sec. \ref{sec:parquet results} we will discuss the divergences of the irreducible vertex in the PAM. In Sec. \ref{sec:FL results} we will evaluate the Fermi liquid parameters in both Hubbard and Anderson lattices. Now, let us begin.

\section{Formalism}\label{sec:Formalism}

In this section we describe the lattice models we will investigate and outline the DMFT method for their solution. Then, we present the decomposition of the Full vertex into irreducible and reducible parts, describe how the irreducible vertex is computed using the Bethe-Salpeter equation (BSE), and show how the irreducible vertex is used to compute the response functions of the lattice. Finally, we decompose the bubble and full vertex into coherent and incoherent parts in order to derive a quasiparticle vertex and evaluate the Fermi liquid parameters.

\subsection{Lattice Models in DMFT}

 Let us begin with a brief description of the Hubbard model and its solution within DMFT before moving on to the Anderson lattice and its solution. 

\subsubsection{Hubbard lattice}\label{sec:Hubbard Model}

The one-band Hubbard model is characterized by the Hamiltonian
\begin{align}
H_{\mathrm{Hubbard}}=
-t\sum_{\langle ij\rangle \sigma} f^\dagger_{i\sigma}f_{i\sigma} +U\sum_i n_{i\uparrow}n_{i\downarrow}
\end{align}
where $t$ is the hopping interaction between nearest neighbors, $f^\dagger_{i\sigma}$ and $f_{i\sigma}$ are the annihilation and creation operator for an electron on site $i$ with spin $\sigma$, and $n_{i\sigma}=f^\dagger_{i\sigma}f_{i\sigma}$ is the associated number operator. (The Anderson lattice, which has interacting states $f$ and non-interacting states $c$, motivates our use of $f$ and $f^\dagger$ to represent the creation and annihilation operators in the Hubbard lattice.)  Here we consider the square-lattice in the paramagnetic phase, and we use the bandwidth of the non-interacting DOS, $D=4t$, as an energy scale, so that our results can be directly compared to those presented in Ref. [{\citen{Schafer2013}}]. We work at half-filling where particle-hole symmetry holds, i.e., at $\mu=U/2$.

In DMFT, one self-consistently maps the lattice problem onto a local Anderson impurity model (AIM) hybridized with a non-interacting bath.\cite{Kotliar1996} The self-consistency condition requires that 
\begin{align}
\sum_{\bm{k}}G(\bm{k},i\nu_n) = G(i\nu_n),
\end{align}
where $G(\bm{k},i\nu_n)$ is the Green's function of the lattice, $G(i\nu_n)$ is the Green's function of the impurity, $i\nu_n$ is a fermionic Matsubara frequency, and we have dropped the spin subscript for brevity. (Recall that we are working with the paramagnetic solutions.) These Green's functions are computed using the Dyson equation
\begin{align}
G(\bm{k},i\nu_n)^{-1} = i\nu_n+\mu - \epsilon_{\bm{k}} -\Sigma_{\bm{k}}(i\nu_n)\\
G(i\nu_n)^{-1} = \mathcal{G}(i\nu_n)^{-1} -\Sigma(i\nu_n),\label{eq:dyson}
\end{align}
where $\mu$ is the chemical potential, $\epsilon_{\bm{k}}$ is the Fourier transform of the hopping matrix, $\Sigma_{\bm{k}}(i\nu_n)$ and $\Sigma(i\nu_n)$ are the self-energy of the lattice and impurity, and $\mathcal{G}(i\nu_n)$ is the Weiss field which, along with the Hamiltonian, defines the AIM. In the limit of infinite coordination, the self-energy is purely local, $\Sigma_{\bm{k}}=\Sigma$, and these equations define a self-consistent method\cite{Kotliar1996}. In DMFT we maintain this result even for finite dimension. 

The local Green's function, $G(\tau)=\langle T_\tau f(\tau)f^\dagger(0)\rangle$, and its Fourier components $G(i\nu_n$) may be measured for a given Weiss field and local Hamiltonian using an impurity solver. Here the local Green's function is computed using the a continuous-time quantum Monte Carlo (CTQMC) algorithm based on the hybridization expansion\cite{Gull2011} solver using the worm algorithm\cite{Gunacker2015} and improved estimators\cite{Hafermann2012,Gunacker2016} to compute these Fourier components. The details of our implementation will be discussed in a subsequent paper\cite{Semon2020}. Analytical continuation to the real frequency domain is accomplished using Pade approximants,\cite{Vidberg1977} so that we can also investigate the spectral functions, $A_i(\nu)=-1/\pi \mathrm{Im}[G_i(\nu)]$.\footnote{The first 500 fermionic frequencies are used for the analytical continuation of the Green's function.}

\subsubsection{Anderson lattice }\label{sec:model}

The Anderson Lattice is characterized by the Hamiltonian\cite{Kotliar1996}
\begin{align}
H_{\mathrm{PAM}}=
\sum_{\bm{k}\sigma}\epsilon^{\vphantom{\dagger}}_{\bm{k}}c^\dagger_{\bm{k}\sigma} c^{\vphantom{\dagger}}_{\bm{k}\sigma}
+\epsilon_f\sum_{\sigma} f^\dagger_{\sigma} f^{\vphantom{\dagger}}_{\sigma}
+U n_{f,\uparrow}n_{f,\downarrow}
+\sum_{\bm{k}\sigma} V^{\vphantom{\dagger}}_{\bm{k}}(c^\dagger_{\bm{k}\sigma} f^{\vphantom{\dagger}}_{\sigma} + f^\dagger_{\sigma} c^{\vphantom{\dagger}}_{\bm{k}\sigma}),
\end{align}
where $c^\dagger_{\bm{k}\sigma}$ and $c_{\bm{k}\sigma}$ are the annihilation and creation operators for the non-interacting lattice state with wavevector $\bm{k}$, spin $\sigma$, and non-interacting energy $\epsilon_{\bm{k}}$, and $f^\dagger_{\sigma}$ and $f_{\sigma}$ are the annihilation and creation operators for the interacting impurity state with spin $\sigma$ and energy $\epsilon_f$. Note that we have absorbed the chemical potential into $\epsilon_f$, which we set to $-U/2$ to ensure particle-hole symmetry.  
We model lattices with a constant interaction parameter $V_{\bm{k}}=V$ and use $V/2=1$ as our energy scale.

In the Anderson lattice, the AIM is supplemented by the self-consistency condition of DMFT: $\sum_{\bm{k}}G_f(\bm{k},i\nu_n) = G_f(i\nu_n)$, where $G_f(\bm{k},i\nu_n)$ is the dressed Green's function and $G_f(i\nu_n)$ is the local Green's function of the $f$-state. In the PAM, the dressed Green's functions of the $c$ and $f$ states are\cite{Kotliar1996,Logan2016}
\begin{align}
G_c(\bm{k},i\nu_n)^{-1} &= i\nu_n - \epsilon_{\bm{k}} - \frac{V^2}{i\nu_n-\epsilon_f-\Sigma_f(i\nu_n)}\\
G_f(\bm{k},i\nu_n)^{-1} &= i\nu_n - \epsilon_f - \Sigma_f(i\nu_n)-\frac{V^2}{i\nu_n-\epsilon_{\bm{k}}},
\end{align}
and the local self energy of the $f$-state, $\Sigma_f(i\nu_n)$, is computed using the Dyson equation, Eq. \ref{eq:dyson}. The impurity solver provides $G_f(i\nu_n)$, and we can solve the PAM using the same DMFT method described above (Sec. \ref{sec:Hubbard Model}).

In this study, we discuss a toy Anderson lattice with a flat non-interacting density of states, $D(\epsilon)=1/W$, with half-bandwidth $W=10$.  This parameterization matches the AIM investigated by Chalupa \textit{et al}. \cite{Chalupa2018}, allowing for a direct comparison. We will also discuss the square Anderson lattice so that we can explore the behavior of the Fermi liquid throughout the Brillouin zone. For the square lattice, we include the nearest-neighbor interactions with hopping $t=2.885$, so that the non-interacting DOS has the same standard deviation as it does in our toy model.   At the one-particle level, the square lattice behaves qualitatively the same as the toy lattice. At the two-particle level, it exhibits qualitatively identical divergences. However, the location of the divergences and phase diagram lines are located at different $T$ and $U$.

With the lattice models described, let us discuss the irreducible vertex. 

\subsection{The two particle irreducible vertex}
\label{sec:Irreducible Vertex Theory}

Just as the local one-particle irreducible vertex (the self-energy) provides the means by which we compute the one-particle quantities of the lattice, the local two-particle irreducible (2PI) vertex provides the means by which we compute the two-particle quantities of the lattice, e.g., the susceptibilities. In DMFT, the 2PI vertex is not incorporated into the self-consistency loop. Instead, it is measured \textit{post hoc} so that one can examine the two-particle response of a given lattice without wasting computational resources. In this section, we define the 2PI vertex and its use.

\begin{table}
\footnotesize
\caption{Nomenclature and notation.}
\label{nomenclature}
\begin{tabular}{ c p{6cm}  c p{6cm}}
\textbf{Symbol} & \textbf{Description} &
\textbf{Symbol} & \textbf{Description}\\
$\nu$ & Fermionic frequency & 
$\omega$ & Bosonic frequency \\
$\bm{k}$ & Wavevector (lattice) & 
$\bm{q}$ & Wavevector (transfer)\\
$q$ & Combined vector $\{\bm{q},\omega\}$ & 
$\mathcal{G}_0$ & Weiss field \\ 
$G(\nu)$ & Local Green's function & 
$G(\bm{k},\nu)$ & Lattice Green's function\\
$\chi_{loc}$ & The local susceptibility & 
$\chi$ &  The lattice susceptibility\\
$\chi_{vc,loc}$ &  Vertex-connected part of the local susceptibility & 
$\chi_{vc}$ &  Vertex-connected part of the lattice susceptibility\\
$\bm{\chi}^\omega_{vc,loc}$ & Local susceptibility matrix  & 
$\bm{\chi}^q_{vc}$ & Lattice susceptibility matrix \\
$\chi_{qp}$ & Coherent part of $\chi_{vc}$ &
$\chi_{inc}$ & Incoherent part of $\chi_{vc}$\\
$\chi_0$ &  Bubble of the $f$ state &
$\chi_{0,qp}$ & Coherent part of $\chi_0$\\
$\chi_{0,inc}$ & Incoherent part of $\chi_0$ &
$A_0$ & Fermi liquid parameter ($\bm{q}=0$ $\omega\rightarrow0$) \\
$F$ & Full vertex & 
$F_0$ &  Fermi liquid parameter ($\omega=0$ $\bm{q}\rightarrow0$)\\
$\Gamma$ & Irreducible vertex &
$\Phi$ & Reducible vertex\\
$\Gamma_{qp}$ &  Coherent vertex  &
$\Phi_{inc}$ & Incoherent vertex \\ 
\multicolumn{4}{c}{\textbf{Superscripts}}\\
$(c)$ & Charge channel  &
$(m)$ & Magnetic (spin) channel \\
&\multicolumn{3}{l}{\textbf{Note:} We discuss the charge channel unless otherwise specified. }\\
$(s)$ & Symmetric part &
$(a)$ & Antisymmetric part \\
\multicolumn{4}{c}{\textbf{Subscripts}}\\
$loc$ & Local quantity & $\sigma$ & spin \\
$c$ & $c$-state (PAM) & 
$f$ & $f$-state \\
&\multicolumn{3}{l}{\textbf{Note:} In the PAM, we discuss the $f$ state quantities unless otherwise specified. }
\end{tabular}
\end{table}

In order to define the irreducible vertex, we first define the susceptibility from which the vertex is computed.  
In the PAM, the local susceptibility of the impurity is given by
\begin{align}
\chi_{loc,f,\sigma\sigma'}(\tau_1,\tau_2,\tau_3,\tau_4)&=\langle T_\tau 
f_\sigma^\dagger(\tau_1) 
f_\sigma^{\vphantom{\dagger}}(\tau_2) 
f_{\sigma'}^\dagger(\tau_3)
f_{\sigma'}^{\vphantom{\dagger}}(\tau_4)\rangle 
- \langle T_\tau f^\dagger_\sigma (\tau_1) f^{\vphantom{\dagger}}_\sigma  (\tau_2)\rangle
  \langle T_\tau f^\dagger_{\sigma'} (\tau_3) f^{\vphantom{\dagger}}_{\sigma'} (\tau_4)\rangle.
\end{align}
Due to time-translation invariance, this object can be described in some representation $l$ (particle-particle, $pp$, particle-hole, $ph$, or transverse particle-hole $\overline{ph}$)
via three Fourier components: $\chi_{loc,f,\sigma,\sigma}(i\nu_n, i\nu_{n'}, i\omega_m)$, where $i\omega_m$ is a bosonic Matsubara frequency. Here we measure the Fourier components in the particle-hole representation, where frequencies $i\nu_n$, $i\nu_{n'}$, and  $i\omega_m$ are associated with the time differences $\tau_1-\tau_2$, $\tau_3-\tau_4$, and $\tau_3-\tau_2$, respectively. Note that $\omega$ denotes a bosonic frequency and $\nu$ a fermionic frequency.  In a single-band impurity model,  the susceptibility of the $f$ states in the charge ($c$) or magnetic/spin ($m$) channels may be easily computed as
\begin{align}
\chi^{(c/m)}_{loc,f}(i\nu_n, i\nu_{n'}, i\omega_m) = \chi_{loc,f,\uparrow\uparrow}(i\nu_n, i\nu_{n'}, i\omega_m)\pm\chi_{loc,f,\downarrow\uparrow}(i\nu_n, i\nu_{n'}, i\omega_m).
\end{align}

Let us condense our notation by making a few observations. First, divergences in the irreducible vertex do not occur in the spin channel.\cite{Schafer2013, Schafer2016, Gunnarson2017, Chalupa2018} Second, correlation in the conduction states occurs through hybridization with impurity states; therefore, all divergences which appear in the conduction state two-particle objects will appear in the $f$-state vertex functions. Third, the equations we will write are diagonal in $\omega$ within the particle-hole representation. 

From these observations, let us assume we are discussing the charge channel of the $f$ states and drop the associated specifiers $f$ and $(c)$. (Note that in the equations derived in this section and in Sec. \ref{sec:FL vertex}, the results hold for both spin and charge channels.) Additionally, let us represent all three frequency objects as matrices with elements given by the fermionic frequencies. For example, the object $\chi^l_{loc,f,c/m}(i\nu_n, i\nu_{n'},\omega)$ is more compactly written as the matrix $\bm{\chi}^\omega_{loc}$ with matrix elements 
$\chi_{loc,nn'}^\omega=\chi_{loc}(i\nu_n, i\nu_{n'}, \omega)$. 
Non-local quantities also depend upon the momentum transfer, $\bm{q}$. Let us absorb this into our notation by using a vector $q=\{\bm{q},\omega\}$, and denote non-local matrix objects as $\bm{\chi}^q$, where this matrix has elements $\chi_{nn'}^q=\chi(\bm{q},i\nu_n, i\nu_{n'}, \omega)$. When appropriate, we will first give the equation for the local impurity quantity and follow it with the equation for the lattice quantity. 
This notation and the nomenclature used hereafter are summarized in Table \ref{nomenclature} for easy reference. 

Now, let us derive the irreducible vertex and its use.

The full vertex is the vertex-connected part of the susceptibility with the outer Green's lines truncated. That is, 
\begin{align}
&\bm{\chi}^{\omega}_{vc,loc} = \bm{\chi}_{loc,0}^{\omega} \bm{F}^{\omega}_{loc}\bm{\chi}_{loc,0}^{\omega}\\
&\bm{\chi}^{q}_{vc}  = \bm{\chi}_{0}^{q} \bm{F}^{q}\bm{\chi}_{0}^{q},
\label{eq:full vertex}
\end{align}
where $\bm{\chi}_{loc,0}^\omega$ and $\bm{\chi}_0^q$ are the bubble diagrams of the impurity and lattice. They have matrix elements
\begin{align}
&\chi_{loc,0}(i\nu_n,i\nu_{n'},i\omega_m)=-\beta G(i\nu_n)G(i\nu_{n'}+i\omega_m)\delta_{nn'}\\
&\chi_{0}(\bm{q},i\nu_n,i\nu_{n'},i\omega)=-\beta\sum_{\bm{k}} G(\bm{k},i\nu_n)G(\bm{k}+\bm{q},i\nu_{n'}+i\omega_m)\delta_{nn'}.
\end{align}
Note that $G$ is the Green's function of the $f$ state, $G_f$, per our simplified notation discussed above.

\begin{figure}
\includegraphics[scale=1]{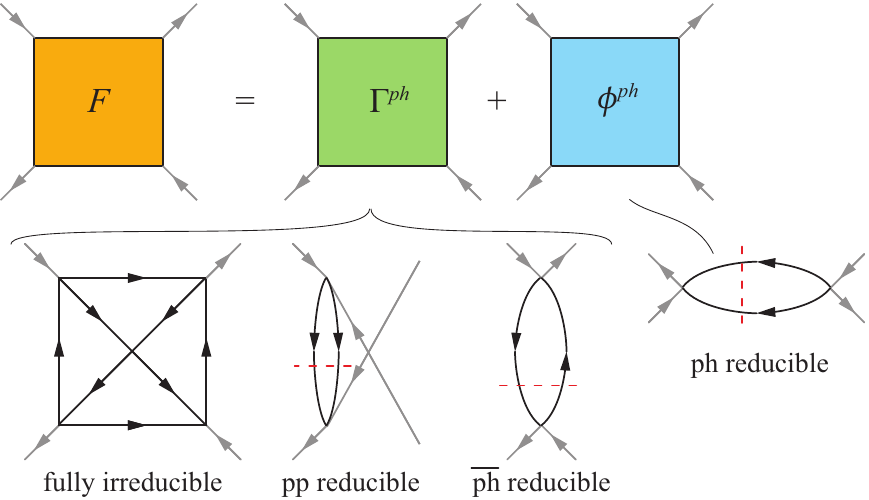}
\caption{The full vertex decomposed into the two-particle irreducible and reducible diagrams in the particle-hole channel. The irreducible diagrams combine both the fully two-particle irreducible diagrams (like the pictured envelope diagram) and the two-particle reducible diagrams in the particle-particle and transverse particle-hole sectors.}
\label{fig:parquet}
\end{figure}

The full vertex, $F$, is decomposed into the two-particle irreducible and reducible diagrams, $\Gamma^l$ and $\Phi^l$, for the particle-hole, $ph$, transverse particle-hole, $\overline{ph}$, and particle-particle channels, $pp$.  
That is,
\begin{align}
F=\Gamma^l + \Phi^l,
\end{align}
for $l \in \{ph, pp, \overline{ph}\}$, as depicted in Fig. \ref{fig:parquet} for the particle-hole channel ($l=ph$). Here, two-particle reducible means that a diagram can be separated by removing two Green's function lines. Again, we will work in the particle-hole representation and drop this superscript for brevity. 

Within this decomposition, one can express the full vertex as the irreducible vertex expanded in ladders of the bubble, i.e.,
\begin{align}
&\bm{F}^{\omega}_{loc}=\bm{\Gamma}^{\omega}_{loc}+\bm{\Gamma}^{\omega}_{loc}\bm{\chi}_{loc,0}^{\omega}\bm{F}^{\omega}_{loc} \\
&\bm{F}^{q}=\bm{\Gamma}^{q}+\bm{\Gamma}^{q}\bm{\chi}_{0}^{q}\bm{F}^{q},
\end{align}
With some algebra, this allows us to relate the irreducible and full vertex:
\begin{align}
&\bm{F}^{\omega}_{loc}=[(\Gamma^{\omega}_{loc})^{-1}-\bm{\chi}_{loc,0}^{\omega}]^{-1}\label{eq:loc:irreducible vertex to full vertex} \\
&\bm{F}^{q}=[(\Gamma^{q})^{-1}-\bm{\chi}_{0}^{q}]^{-1},\label{eq:latt:irreducible vertex to full vertex} 
\end{align}
where we emphasize that the $(\cdots)^{-1}$ notation indicates the matrix inverse of the matrix $(\cdots)$. Then, the susceptibility and irreducible vertex are related through the BSE as
\begin{align}
&(\bm{\chi}^{\omega}_{loc})^{-1} = (\bm{\chi}_{loc,0}^{\omega})^{-1} - \bm{\Gamma}_{loc}^{\omega}\label{eq:loc:irreducible vertex}\\
&(\bm{\chi}^{q})^{-1} = (\bm{\chi}_0^{q})^{-1} - \bm{\Gamma}^{q}\label{eq:latt:irreducible vertex}
\end{align}
As shown in Fig. \ref{fig:vertex-connected parts}, the total susceptibility is the combination of bubble and vertex-connected parts.

Finally, we compute the single-time susceptibility of the lattice, $\chi(\bm{q},\omega)$, an observable. It is computed by summing all elements in the matrix $\bm{\chi}^q$.\footnote{We keep the lowest $200\times200$ fermionic components in this summation, truncating the remaining high-frequency components. No analytical continuation is used for the susceptibilities, so only the first few Bosonic components are measured.} That is,
\begin{align}
\chi(\bm{q},\omega) = \sum_{nn'}\chi(\bm{q},i\nu_n,i\nu_{n'},\omega).
\end{align}
The single-time vertex-connected susceptibility is computed in the same manner.

In a typical DMFT calculation of the lattice susceptibility, the local susceptibilities of the impurity are measured by the impurity solver. Then, the local charge and spin susceptibilities are computed. Next, the irreducible vertex function is computed from Eq. (\ref{eq:loc:irreducible vertex}). To proceed, one assumes that that the irreducible vertex function is purely local, i.e., $\bm{\Gamma}^q = \bm{\Gamma}^\omega_{loc}$, and computes the lattice susceptibility using Eq. (\ref{eq:latt:irreducible vertex}). (This is equivalent to the DMFT approximation of the one-particle irreducible vertex, the self-energy, as purely local: $\Sigma_{\bm{k}}=\Sigma$.)  As an aside, one can avoid the numerical issues associated with irreducible vertex and its divergences by instead computing the local full vertex, then the non-local full vertex, and finally the lattice susceptibility as suggested in Refs. \citen{Galler2017,Rohringer2018}. However, the conceptual concerns remain. 

\subsection{The quasiparticle vertex}
\label{sec:FL vertex}

Here we present an alternative decomposition of the full vertex into coherent and incoherent parts rather than reducible and irreducible parts. To do this, we focus on irreducibility  with respect to the quasiparticle part of the electron Green's function instead of the full Green's function. 

Let us begin by splitting the local and lattice bubble diagrams, $\bm{\chi}_{loc,0}^{\omega} $ and $\bm{\chi}_0^q$, into coherent (quasiparticle), $\bm{\chi}_{loc,0,qp}^\omega$ and $\bm{\chi}_{0,qp}^q$, and incoherent, $\bm{\chi}_{loc,0,inc}^\omega$ and $\bm{\chi}_{0,inc}^q$, parts, i.e.,
\begin{align}
&\bm{\chi}^{\omega}_{loc,0} = \bm{\chi}^{\omega}_{loc,0,qp} + \bm{\chi}_{loc,0,inc}^{\omega}.\\
&\bm{\chi}^{q}_0 = \bm{\chi}^{q}_{0,qp} + \bm{\chi}_{0,inc}^{q}.
\end{align}
Here, we define the coherent bubble as the low-frequency region which avoids the branch cuts in the Green's functions. That is, $G(i\nu_n < 0)G(i\nu_n+\omega_n > 0)$ or
\begin{align}
&\bm{\chi}_{loc,0,qp}^{\omega} =
 \left\{ \begin{array}{ll} 
                \bm{\chi}_{loc,0}^{\omega} & \hspace{5mm} \nu \in (-\omega,0)\\
                0 & \mathrm{otherwise}\\
\end{array} \right. \\
&\bm{\chi}_{0,qp}^{q} =
 \left\{ \begin{array}{ll} 
                \bm{\chi}_{0}^{q} & \hspace{5mm} \nu \in (-\omega,0)\\
                0 & \mathrm{otherwise}\\
\end{array} \right.
\label{eq:qp bubble}
\end{align}

This is a reasonable definition, for a few reasons. Qualitatively, Fermi liquid theory is a low-energy theory focused on the behavior of quasiparticles on or very near the Fermi surface. Our definition is limited to this low-energy regime. More quantitatively, in the Green's function approach to Landau Fermi liquid theory, one describes the quasiparticles via the pole in the Green's function which has some quasiparticle weight $z_{\bm{k}}$ and renormalized energy $\epsilon_{\bm{k}}$. At finite temperature or in a system with impurities, these quasiparticles scatter (at rate $\gamma$) which disperses these poles. Still, if the quasiparticle lifetime remains long, $\gamma\rightarrow0$, and if the energy and momentum exchange between quasiparticles vanishes, $\omega\rightarrow0$ and $\bm{q}\rightarrow0$, then we can capture the quasiparticle bubble in Landau Fermi liquid theory as\cite{Nozieres1966}
\begin{align}
\chi_{0,qp}^{\bm{q}}(\omega) = \sum_{\bm{k}}z_{\bm{k}}^2\frac{2i\pi\delta(\epsilon_{\bm{k}}-\mu)}{2i\gamma-\omega-\bm{v}_{\bm{k}}\cdot\bm{q}},\label{eq:coherent bubble}
\end{align}
where $z_{\bm{k}}=(1+\partial\Sigma_{\bm{k}}/\partial\omega|_{\omega=\mu})^{-1}$, $\mu$ is the chemical potential, and $\bm{v}_{\bm{k}}=\partial\epsilon_{\bm{k}}/\partial{\bm{k}}$ is the quasiparticle velocity. This definition also appears in the theory of disordered Fermi Liquids\cite{Nozieres1966,Castellani1987}.

From this equation, we can see that the primary information captured by the coherent part of the bubble (the quasiparticle bubble) is contained between and near the two poles, i.e., in the low-frequency domain, particularly when the quasiparticle is well defined ($\gamma\rightarrow0$). Note that the remaining incoherent part must primarily capture the high frequency domain away from the poles. Thus, our definition which takes all information between and at the two poles, should provide a reasonable definition of the coherent bubble.  Furthermore, it is a definition restricted to objects which are naturally computed in finite temperature DMFT.

Next we decompose the full vertex into coherent, $\bm{\Gamma}_{qp}$ and incoherent, $\bm{\Phi}_{inc}$, parts
\begin{align}
&\bm{F}^{\omega}_{loc} = \bm{\Gamma}_{loc,qp}^{\omega} + \bm{\Phi}_{loc,inc}^{\omega}
\label{eq:loc:FL decomposition}\\
&\bm{F}^{q} = \bm{\Gamma}_{qp}^{q} + \bm{\Phi}_{inc}^{q}
\label{eq:latt:FL decomposition}
\end{align}
Then we can expand the full vertex in ladders of the coherent bubble and vertex, i.e.,
\begin{align}
&\bm{F}^{\omega}_{loc} =  \bm{\Gamma}_{loc,qp}^{\omega} +  \bm{\Gamma}_{loc,qp}^{\omega} \bm{\chi}^{\omega}_{loc,0,qp}\bm{F}^{\omega}_{loc}\label{eq:loc:full vertex to FL vertex}\\
&\bm{F}^{q} =  \bm{\Gamma}_{qp}^{q} +  \bm{\Gamma}_{qp}^{q} \bm{\chi}^{q}_{0,qp}\bm{F}^{q}.\label{eq:latt:full vertex to FL vertex}
\end{align}
With some algebra. we can write the coherent vertex in terms of the full vertex as
\begin{align}
&\bm{\Gamma}_{loc,qp}^{\omega} = [(\bm{F}^{\omega}_{loc})^{-1} + \bm{\chi}_{loc,0,qp}^{\omega}]^{-1}\\
&\bm{\Gamma}_{qp}^{q} = [(\bm{F}^{q})^{-1} + \bm{\chi}_{0,qp}^{q}]^{-1}.
\end{align}
Note that if the coherent bubble vanishes, i.e., at $\omega=0$, then the quasiparticle vertex and full vertex are identical. 

\begin{figure}
\includegraphics{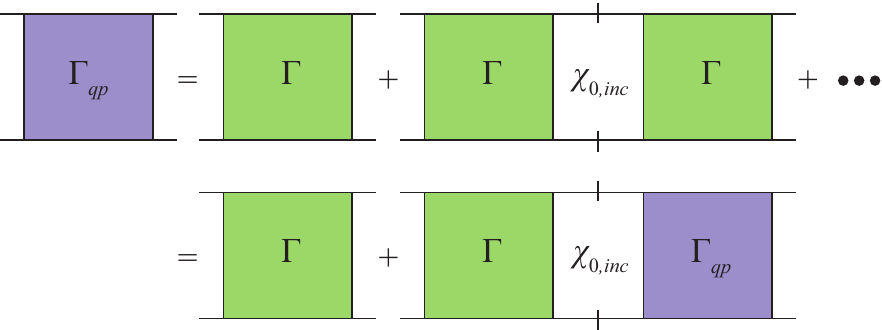}
\caption{Diagrammatic connection between coherent and irreducible vertices.}\label{fig:FL vertex diagram}
\end{figure}

Using Eqs. (\ref{eq:loc:irreducible vertex to full vertex}) and  (\ref{eq:latt:irreducible vertex to full vertex}) we can compare the irreducible and quasiparticle vertices:
\begin{align}
&\bm{\Gamma}_{loc,qp}^{\omega} = \frac{\bm{\Gamma}^{\omega}_{loc}}{\bm{1} - \bm{\chi}_{loc,0,inc}^{\omega}\bm{\Gamma}^{\omega}_{loc}},\label{eq:loc:irreducible vertex to FL vertex}\\
&\bm{\Gamma}_{qp}^{q} = \frac{\bm{\Gamma}^{q}}{\bm{1} - \bm{\chi}_{0,inc}^{q}\bm{\Gamma}^{q}},\label{eq:latt:irreducible vertex to FL vertex}
\end{align}
where the denominator represents a matrix inverse. From this result, we can see that the coherent vertex will not diverge when the irreducible vertex diverges, unless $\bm{\chi}_{inc}^{q}=\bm{0}$. Additionally, we see that the coherent vertex is an expansion of the irreducible vertex in ladders of the incoherent bubble diagrams, as shown in Fig. \ref{fig:FL vertex diagram}. In effect, the incoherent part of the bubble screens the divergences which arise in the irreducible vertex.

Eqs. (\ref{eq:loc:irreducible vertex to full vertex})  and (\ref{eq:latt:irreducible vertex to full vertex}) also allow us to relate the impurity and lattice quantities when we assume the irreducible vertex is purely local. For example, we can write the full vertex of the lattice in terms of the local full vertex and the non-local bubble\cite{Galler2017} 
\begin{align}
\bm{F}^q = [(\bm{F}_{loc}^\omega)^{-1} - \bm{\chi}_0^{nl,q}]^{-1},\label{eq:Full vertex result}
\end{align}
where $\chi_0^{nl,q} = \chi_0^q-\chi_{loc,0}^\omega$.  This approach allows us to avoid the numerical issues associated with the divergences of the irreducible vertex.

Finally, let us show how one can extract the Fermi liquid parameters from the quasiparticle vertex. 

\subsection{The Fermi liquid parameters}
\label{sec:fl parameters}

In Landau Fermi-liquid theory, the free energy is expressed as\cite{Nozieres}
\begin{align}
F=\sum_{\bm{k}\sigma}(\epsilon_{\bm{k}}-\mu)\delta n_{\bm{k}\sigma}
+\sum_{\bm{k}\sigma,\bm{k}'\sigma'}f_{\bm{k}\sigma,\bm{k}'\sigma'}\delta n_{\bm{k}\sigma}\delta n_{\bm{k'}\sigma'},
\end{align}
where $\epsilon_{\bm{k}}$ is the energy of the quasiparticle, $f_{\bm{k}\sigma,\bm{k}'\sigma'}$ describes the interaction between two quasiparticles, and $\delta n_{\bm{k}\sigma}$ is the difference between the occupation of state $\bm{k},\sigma$ and the Fermi function.  The interaction can be decomposed into symmetric and antisymmetric Fermi liquids parameters and expanded using, e.g., the Legendre polynomials as
\begin{align}
f_{\bm{k}\sigma,\bm{k}'\sigma'}\propto\Sigma_l P_l(cos\theta)(F_l^{(s)}+\sigma\sigma'F_l^{(a)}),
\end{align}
where $P_l$ are the Legendre polynomials and  $\theta$ is the angle between $\bm{k}$ and $\bm{k}'$. In DMFT, we lose the $\bm{k}$ dependence and can only evaluate the zeroth order parameters, $F_0^{(s)}$ and $F_0^{(a)}$. 

Within our formalism, the quasiparticle vertex contains the related $A_0^{(s/a)}$ Fermi liquid parameters:
\begin{align}
A_{0,loc}^{(s/a)} &= z^2 D^*(0)\Gamma_{loc,qp}^{(c/m)}(i\nu_{-1},i\nu_{-1},i\omega_{1}) \\
A_0^{(s/a)} &= z^2 D^*(0)\Gamma_{qp}^{(c/m)}(\bm{q}=0,i\nu_{-1},i\nu_{-1},i\omega_{1}), \label{eq:fermi liquid parameters} 
\end{align}
where $\nu_{-1}=-i\pi T$ and $\omega_1=i2\pi T$. The $F_0^{(s/a)}$ parameters may be easily computed from the $A_0^{(s/a)}$ parameters via\cite{Nozieres1966}
\begin{align}
F_0^{(s/a)} = \frac{A_0^{(s/a)}}{1-A_0^{(s/a)}}.
\end{align}
The $A_0$ parameters are called the static parameters whereas the $F_0$ parameters are called the dynamic parameters.

\textcolor{\changescolor}{
}

\textcolor{\changescolor}{
}

\textcolor{\changescolor}{
}

\textcolor{\changescolor}{
}

\textcolor{\changescolor}{
}

With the formalism established, let us begin discussing our results. We start with the one-particle properties.

\section{One Particle Properties}
\label{sec:PAM results}

\begin{figure}
\includegraphics{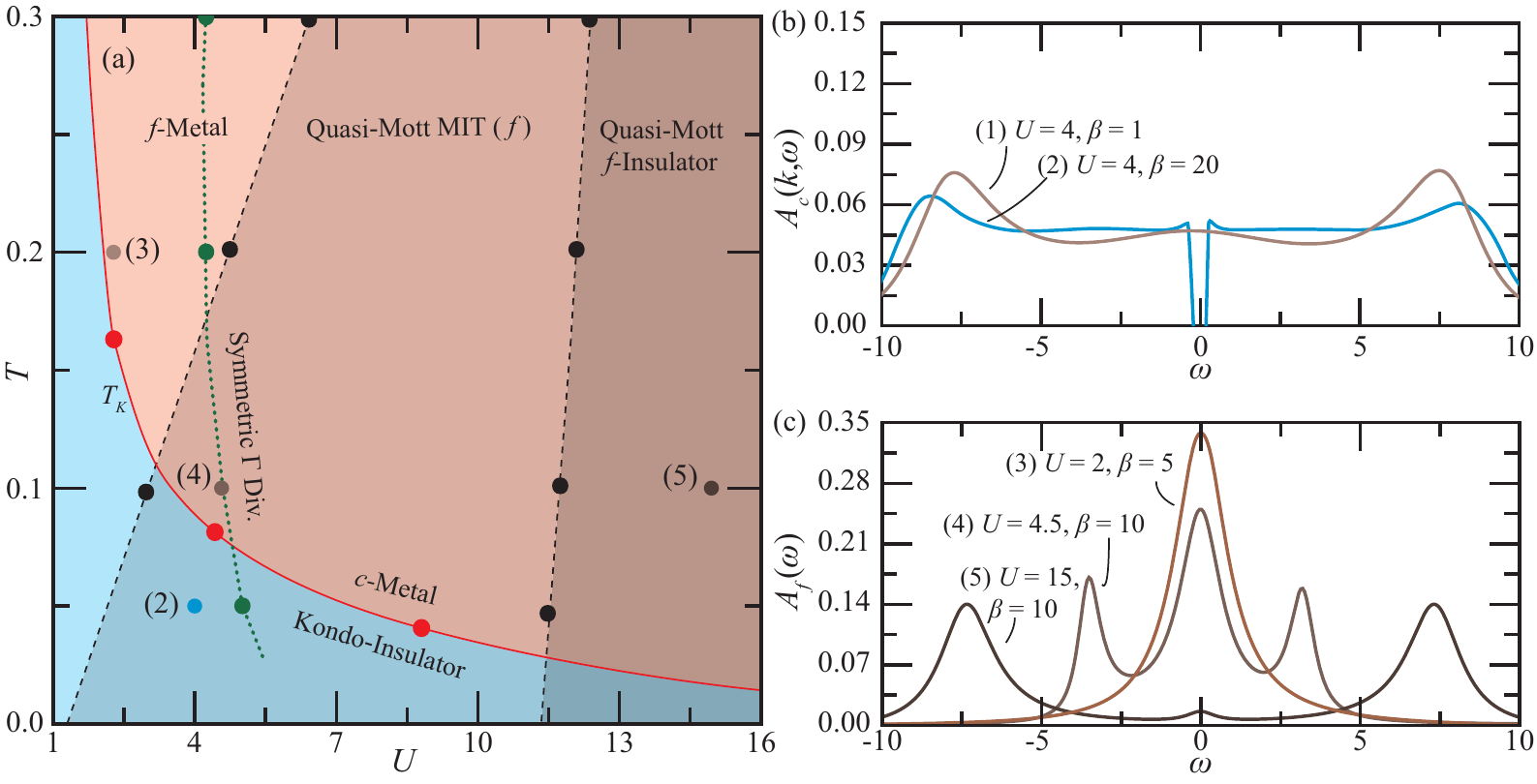}
\caption{ (a) Phase diagram of the PAM and the first symmetric divergence of the irreducible vertex function, $\Gamma$, (b) the spectral function of the conduction states in the Kondo insulating (blue) and metallic (orange) states, and (c) the $f$-state spectral functions in the metallic state as it transitions from a metallic (light orange) to Mott-insulator-like (dark orange) state. Each curve in (b) and (c) is labeled by a numeral in (a). The Insulating and metallic phases are delineated by the Kondo temperature (red line). While the $f$-state does not undergo an independent phase transition, we mark its transition between metallic and Mott-like regions (dashed black lines). The first symmetric divergence of the irreducible vertex is also shown (dashed green line; see Sec. \ref{sec:parquet results}).  This divergence does not correlate with any phase transition. 
}
\label{fig:phase diagram 1}
\end{figure}

The one-particle properties of both Hubbard\cite{bulla1999,blumer2003,Schafer2013} and Anderson lattices\cite{Medici2005, Lorenzo2008, Logan2016} are well established in the literature. Still, it is useful to provide a few results here which will provide context for the discussions on the irreducible and quasiparticle vertices which follow. In particular, it will be useful to understand the phase diagram of the PAM. (The half-filled Hubbard model has a less complicated diagram, featuring a single transition from a metal to a Mott insulator at low temperatures.)  Here, we discuss the toy Anderson lattice described in Sec. \ref{sec:model}. (Recall that the square lattice will have qualitatively identical one-particle properties.) Figure \ref{fig:phase diagram 1}(a) provides the phase diagram of this material. Let us discuss.

At high temperatures the conduction states do not interact with the impurity, and the spectral function of the conduction states, $A_c(\omega)$, approaches that of the non-interaction density of states, as shown in Fig. \ref{fig:phase diagram 1}(b). That is, in the atomic-limit the PAM is a $c$-metal. As the temperature approaches the Kondo temperature, $T_K$, the hybridization between impurity and conduction states becomes important. This hybridization suppresses the spectral functions of both conduction and $f$ states, $A_c(\omega)$ and $A_f(\omega)$, at the Fermi level.
As the temperature falls below the Kondo temperature, this hybridization creates a band  gap at the Fermi level, and the PAM becomes a Kondo insulator, as shown in Fig. \ref{fig:phase diagram 1}(a) and (b).  

While the PAM at half-filling does not undergo a Mott-Hubbard MIT, the $f$ states do exhibit a transition reminiscent of such a MIT, as shown in Fig. \ref{fig:phase diagram 1}(a) and (c), and as discussed in the literature\cite{Medici2005, Lorenzo2008,Logan2016}. That is, as the interaction strength increases, the spectral function of the $f$ states transitions from a single-peaked, Gaussian-like form (metal), to a triple-peaked form (two hubbard bands emerge at $\pm U/2$), and finally to a double-peaked form (Mott insulator). However, some small spectral weight remains between these two peaks and at the Fermi-level, so that the $f$-states never become truly insulating. Moreover, the conduction states hybridize less to this quasi-Mott insulating $f$ state than they do to the metallic $f$ state, as there is less spectral weight at the Fermi level. Thus, not only do the conduction states not undergo a Mott-like transition, but the Kondo temperature also decreases as $U$ increases. 

For any $T$ and $U$, both conduction and $f$ states remain half-filled, as guaranteed by the particle-hole symmetry of the PAM Hamiltonian with $\epsilon_f=-U/2$. 

With the well known one-particle behavior of this PAM introduced, let us discuss the two-particle results. (We refer the reader to Refs. [\citen{Medici2005, Lorenzo2008, Logan2016}] for a more thorough discussion of the one-particle behavior of the Anderson lattice.) Let us begin with a discussion of the divergence of the irreducible vertex.

\section{Divergences of the Irreducible Vertex}
\label{sec:parquet results}

In this section, we will focus on the toy Anderson lattice for which we presented one-particle results in Sec. \ref{sec:PAM results}. Divergences of the irreducible vertex in the Hubbard lattice have been well established in the literature.\cite{Schafer2013,Schafer2016,Gunnarson2016} Here we use the Hubbard lattice primarily to investigate the generality of our observations and conclusions.

\begin{figure}
\includegraphics{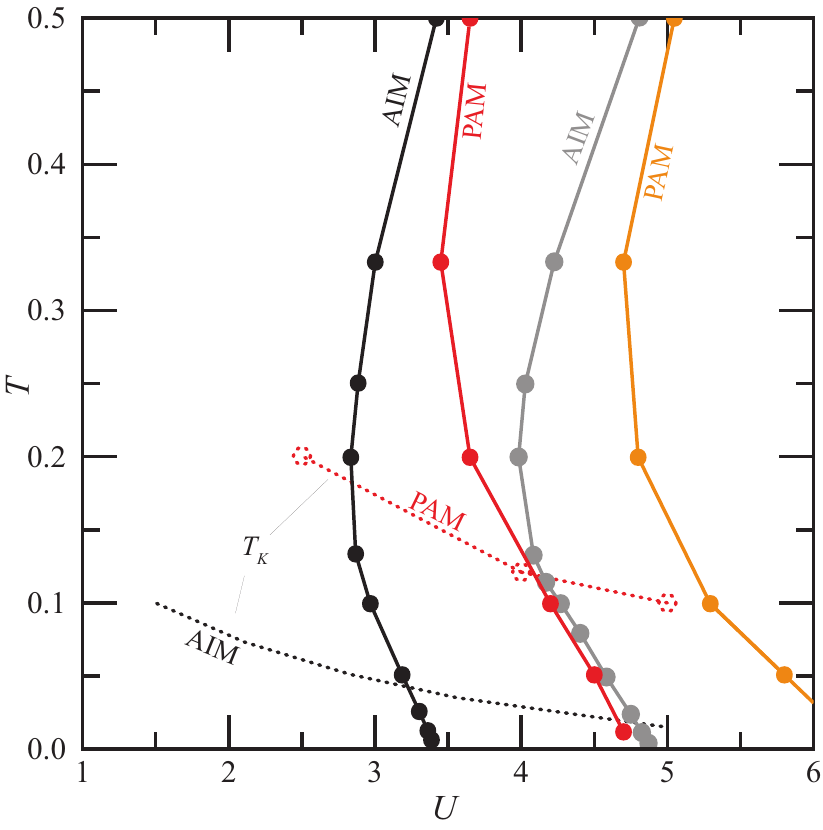}
\caption{ Comparison between $T$-$U$ diagrams of the first anti-symmetric (black, red) and symmetric (gray, orange) divergences line in two models (AIM, PAM). The AIM data is from Ref. [{\citen{Chalupa2018}}].  The Kondo temperature is shown for both models. In the PAM, this delineates the boundary above which the PAM is metallic and below which it is a Kondo insulator. The PAM exhibits both a higher Kondo temperature and also divergence lines which occur at larger interaction strengths than the AIM. }
\label{fig:phase diagram}
\end{figure}

\begin{figure}
\includegraphics{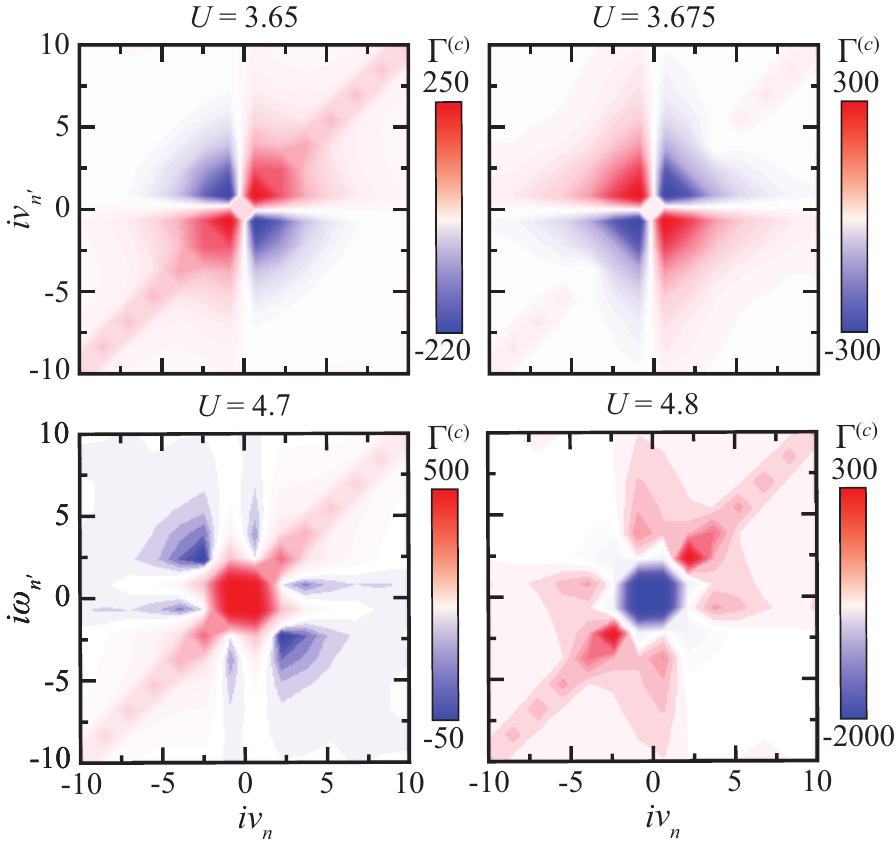}
\caption{ The real part of the static ($\omega=0$) irreducible vertex on either side of the antisymmetric (top row) and symmetric (bottom row) divergence lines at $\beta=5$. To see the divergence, note that blue (red) denote a large and negative (positive) value. Therefore, when one sees that a region of a graph changes colors as one looks from one column to another, the vertex has diverged at those frequencies. For example, examine the first quadrant of $\Gamma_c$ for $U=3.65$ and $3.675$, wherein the sign flips from a large positive to a large negative value (and the color flips from dark red to dark blue).}
\label{fig:irreducible vertex divergence}
\end{figure}

Before we proceed, however, it is helpful to first provide a spectral decomposition of the inverse susceptibility. That is, we write $\chi_{loc}^{-1}$ in terms of its eigenvectors $\overrightarrow{V}_i(i\omega)$ and the eigenvalues $\lambda_i$ of $\chi_{loc}$ as\cite{Springer2019}
\begin{align}
\chi_{loc}(i\nu_n, i\nu_{n'}, i\omega)^{-1} = \sum_i V_{i,n}^*(i\omega) \lambda^{-1}_i V_{i,n'}(i\omega),
\end{align}
where $V_{i,n}(i\omega)$ is element $n$ of the eigenvector $\overrightarrow{V}_i(i\omega)$.
As noted in Ref. [\citen{Schafer2016}], a divergence in $\bm{\Gamma}=\bm{\chi}_{vc}^{-1}-\bm{\chi}_0^{-1}$ occurs when an eigenvalue $\lambda_i$ of the local susceptibility approaches and then crosses zero, such that $\lambda_i^{-1}$ diverges. It has also been noted that the eigenvectors associated with these divergences tend to alternate between antisymmetric and symmetric behavior as one increases the Hubbard interaction strength.\cite{Schafer2016,Chalupa2018} Indeed, one can plot alternating antisymmetric and symmetric divergence lines on a $T$-$U$ diagram for the Hubbard lattice\cite{Schafer2013,Schafer2016}, AIM\cite{Chalupa2018}, and PAM (Fig. \ref{fig:phase diagram}.) For reference, Fig. \ref{fig:irreducible vertex divergence} presents the irreducible vertex on either side of these divergence lines in the PAM.

\begin{figure}
\includegraphics{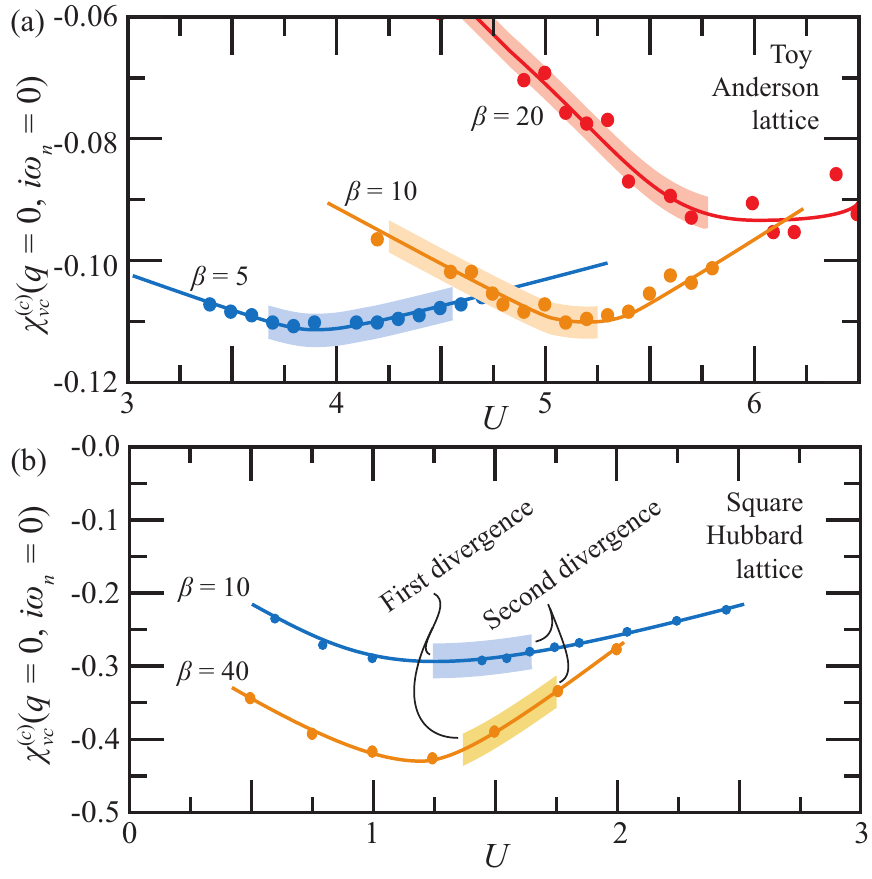}
\caption{The vertex-connected part of the charge susceptibility of the $f$-state near the first divergence lines in (a) the toy Anderson lattice and (b) the square Hubbard lattice at a few temperatures. The first divergences occur near $\partial \chi_{c,f} /\partial U \approx 0$ in both models. The first and second divergence divergence lines are indicated by the onset and termination of the shaded regions behind each curve.  }
\label{fig:susceptibility}
\end{figure}

The irreducible vertex divergences in the PAM behave much like divergences in the Hubbard lattice or Anderson impurity. These divergences are presaged by the appearance of more substantial and negative diagonal components of $\bm{\chi}^\omega_{vc,loc}$, i.e., by the suppression of charge fluctuations. (The first divergence appears shortly before the diagonal becomes negative, while the second divergence occurs as the diagonal becomes negative.\cite{Springer2019})
Moreover, the first divergence lines appear near the maximum magnitude of the (negative) $\chi_{vc}^{(c)}$, as shown in Fig. \ref{fig:susceptibility}. Note that the first divergence should not be connected to a physical response, as it is antisymmetric and its contributions to any observable therefore vanish. In contrast, the second divergence is symmetric and might be physical.\cite{Schafer2013,Schafer2016,Gunnarson2016} The divergences forming in the vicinity of the maximum, negative vertex corrections to the charge susceptibility supports the hypothesis that the suppression of charge fluctuations leads to these divergences.  

While the second divergence in the Hubbard model occurs near the onset  of the Mott MIT\cite{Schafer2013},  the second and symmetric divergence line does not coincide with some phase change  in the PAM, as shown in Fig. \ref{fig:phase diagram 1}.   Indeed, there is no true Mott MIT in the Anderson impurity or lattice, but there are divergences. In general, our results support that these divergences do not foreshadow a phase transition\cite{Chalupa2018}; instead, they are the ubiquitous consequence the Hubbard interaction suppressing charge fluctuations in the strong-coupling regime.\cite{Gunnarson2016}

Figure \ref{fig:phase diagram} also compares the first two divergence lines of the PAM and an AIM model with the same non-interacting bath and coupling. At high temperatures, i.e., as both models approach the atomic limit, these divergence lines converge. This is the expected behavior, as the models are identical in the atomic limit. At low temperatures, however, the divergences in the PAM appear at larger values of $U$. As shown, the Kondo temperature is also substantially larger in the PAM. Therefore, we can conclude that the coupling in the AIM is suppressed by the enhanced hybridization in the PAM, delaying the onset of the strong-coupling regime and its  associated divergences. This aligns with the relative magnitude of their Kondo temperatures, as shown in Fig 3. and as discussed in Ref. [\citen{Rice1986}].

\begin{figure}
\includegraphics{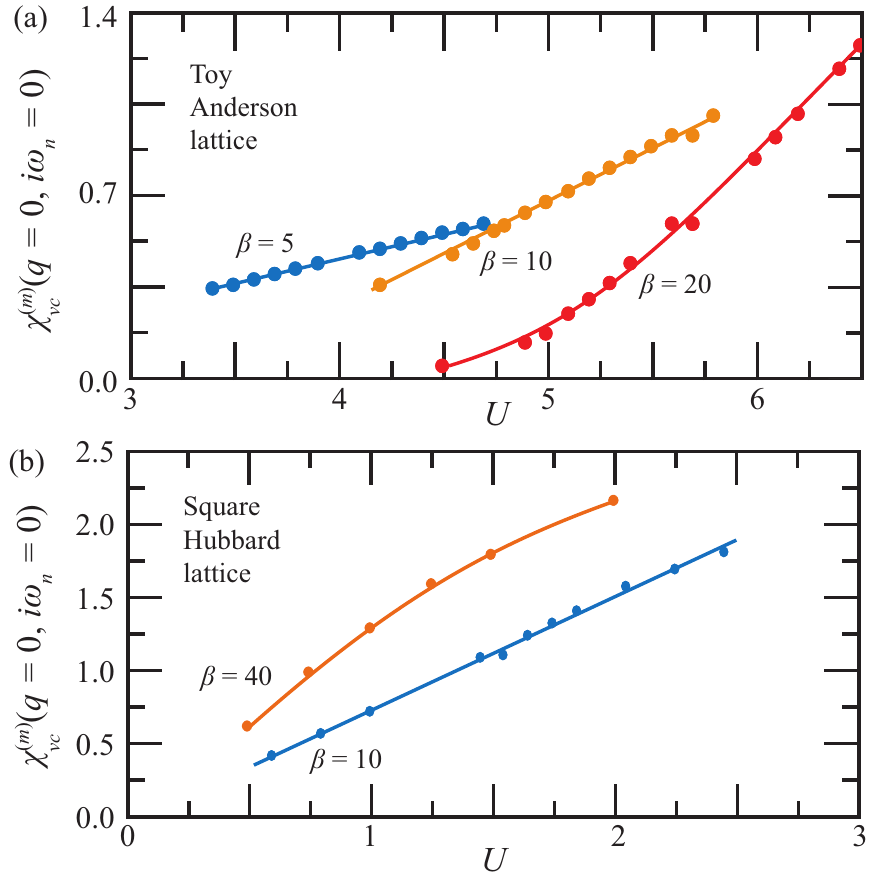}
\caption{The spin susceptibility of the interacting state near the first divergence lines in (a) the toy Anderson lattice and (b) the square Hubbard lattice at a few temperatures. No spin response is connected to the divergences in the charge channel, and no divergences occur in the spin channel.   }
\label{fig:spin susceptibility}
\end{figure}

As expected, there is no divergence in the spin channel. This has been true in every model studied so far.\cite{Schafer2013,Chalupa2018}  Still, the spin susceptibilities as a function of $U$ are presented in Fig. \ref{fig:spin susceptibility} at various temperatures and both lattice models for the interested reader. Now, let us discuss the behavior of the Fermi liquid near the divergences.

\section{Response of the Fermi liquid}
\label{sec:FL results}

Here we investigate the Fermi liquid in the square Anderson and Hubbard lattices. We investigate the square, rather than toy, Anderson lattice so that we can investigate the $\bm{q}\neq0$ behavior of the Fermi liquid, because we are no longer comparing our results with those of Chalupa \textit{et al.}, and because we will be comparing the results to the square Hubbard lattice. As stated in Sec. \ref{sec:model}, our previous observations about the toy Anderson lattice can also be made for the square Anderson lattice. Still, one should note that the divergences occur at larger $U$ and the Kondo temperature is suppressed by the van Hove singularity in the non-interacting DOS. 
Here we will discuss the Fermi liquid parameters, the structure of the Fermi liquid vertices, and the coherent and incoherent part of the lattice susceptibility. Let us begin with a discussion of the Fermi liquid parameters. 

\subsection{Fermi liquid parameters}
\label{sec:fl parameters results}

\begin{figure}
\includegraphics{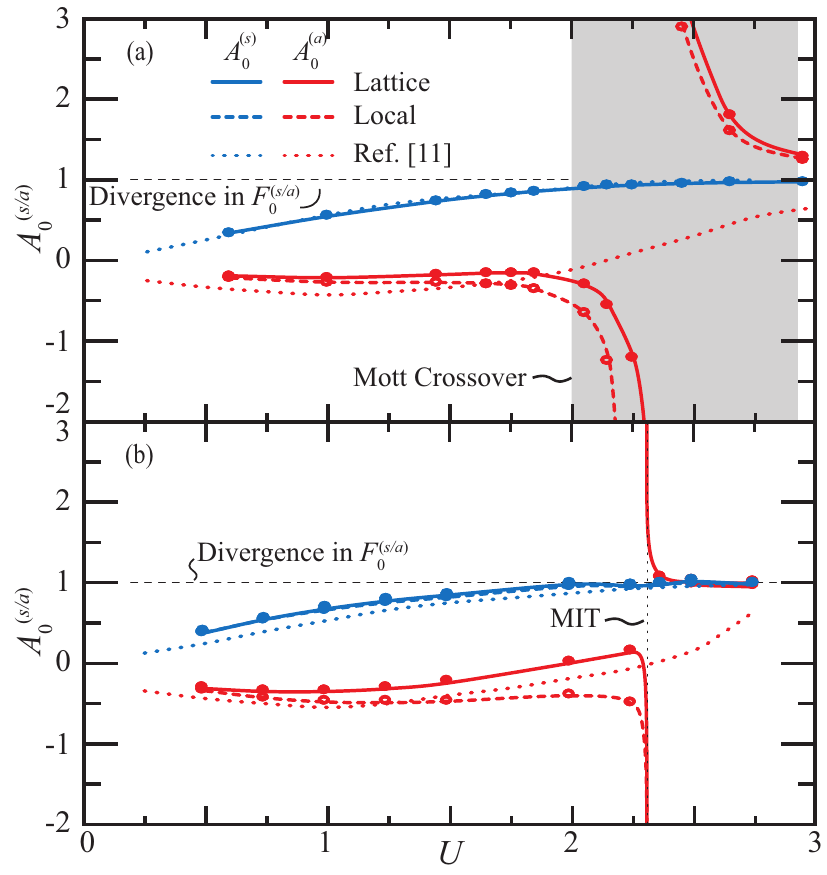}
\caption{\textcolor{\changescolor}{The Fermi liquid $A_0$ parameters in the square Hubbard lattice at (a) $\beta=10$ and (b) $\beta=40$  for variations in $U$. The region of the Mott crossover ($\beta=10$) is indicated by a gray background, while the  Mott  MIT ($\beta=40$)  is indicated by a dotted line. The $A_0=1$ line across which the $F_0$ parameters diverge is shown by a dashed line. The symmetric parameters converge to this line  from below during a Mott transition or crossover. The antisymmetric parameters diverge at the Mott MIT, after which they become  non-physical. (Typically, this sort of divergence indicates a magnetic instability. Here, it indicates that the Fermi-liquid theory breaks down in a Mott regime.) The results of Ref. \citen{Krien2019_fermiliquid} on the triangular lattice are shown for comparison.  }
}
\label{fig:hubbard fl parameters A}
\end{figure}

\begin{figure}
\includegraphics{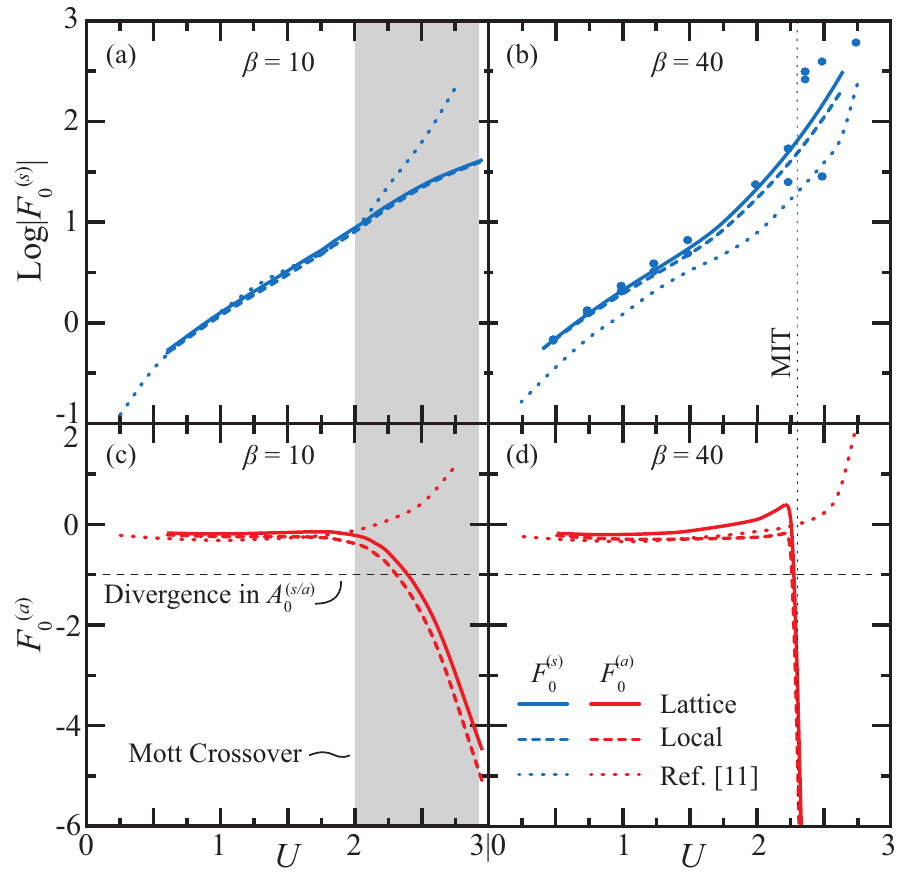}
 \caption{\textcolor{\changescolor}{ The Fermi liquid $F_0$ parameters in the square Hubbard lattice at (a,c) $\beta=10$ and (b,d) $\beta=40$ for variations in $U$. The region of the Mott crossover ($\beta=10$) is indicated by a gray background, while the  Mott  MIT ($\beta=40$)  is indicated by a dotted line. The $F_0=-1$ line across which the $A_0$ parameters diverge is shown by a dashed line. Small errors for $A_0^{(s)}\approx 1$ lead to extremely large errors in $F_0^{(s)}$ and spurious sign changes, motivating our plotting of $|F_0^{(s)}|$ for (a) and (b). $F_0^{(a)}\rightarrow-\infty$ at the Mott MIT, reflecting the breakdown of the theory in the Mott regime. The results of Ref. \citen{Krien2019_fermiliquid} on the triangular lattice are shown for comparison.  }}
\label{fig:hubbard fl parameters F}
\end{figure}

Figures \ref{fig:hubbard fl parameters A} and \ref{fig:hubbard fl parameters F} show the Fermi liquid parameters in the Hubbard lattice for variations in the interaction strength \textcolor{\changescolor}{during a Mott crossover ($\beta=10$) and Mott transition ($\beta=40$). Recall that a crossover indicates a gradual transition from Fermi-liquid to Mott physics, and the transition indicates a sudden change from a Fermi-liquid (metallic) state to an insulating, Mott state. Here, the Mott MIT is found by examining the spectral function. The transition is accompanied by a sharp drop in $D(0)$, when the quasiparticle peak at the Fermi surface vanishes. In contrast, the onset of the crossover region is identified only qualitatively here by the formation of prominent Hubbard bands in the spectral function.
}

\textcolor{\changescolor}{
}


As expected\cite{Nozieres}, the symmetric $A_0$ parameter converges to unity from below, $A_0^{(s)}\rightarrow1$, at the Mott transition or during the Mott crossover. Moreover, the antisymmetric parameters become non-physical when the Mott regime is entered. That is, the Fermi liquid theory  breaks down if there are no well-defined quasiparticles at the Fermi surface. ($A_0^{(a)})\rightarrow\pm\infty$ or $F_0^{(a)}=-1$ can also indicate a magnetic instability. Here, no such magnetic instability exists and this result indicates the \textit{expected} breakdown of the theory rather than some magnetic phase transition.)
In general, one should not analyze the behavior of the Fermi liquid parameters after a Mott transition, as there is no Fermi liquid to analyze. We show the parameters during the crossover and after the transition for completeness, for comparison with Krien \textit{et al.}, and to illustrate the divergence. However, one should not look to extract physical meaning beyond this: The Fermi-liquid no longer exists.


\textcolor{\changescolor}{
As we have discussed, Krien \textit{et al.} also examined the Fermi liquid parameters in the Hubbard model in DMFT. It is important to compare the results, as we have derived a different expression for the Fermi liquid parameter. Before continuing, we note that we are simulating a square lattice, and they examined triangular lattice. Still, we expect that the Fermi liquid parameters should be similar between these two lattice models, as the quantities in question are primarily local.
}

\textcolor{\changescolor}{
Indeed, the two theories predict nearly identical behavior in the Fermi liquid regime. As we enter the Mott regime, however, the predictions differ drastically. As we have discussed, this is not particularly concerning, as no insight should be extracted from the behavior of the Fermi-liquid parameters in the Mott regime. 
}

\textcolor{\changescolor}{
}

\begin{figure}
\includegraphics[scale=1]{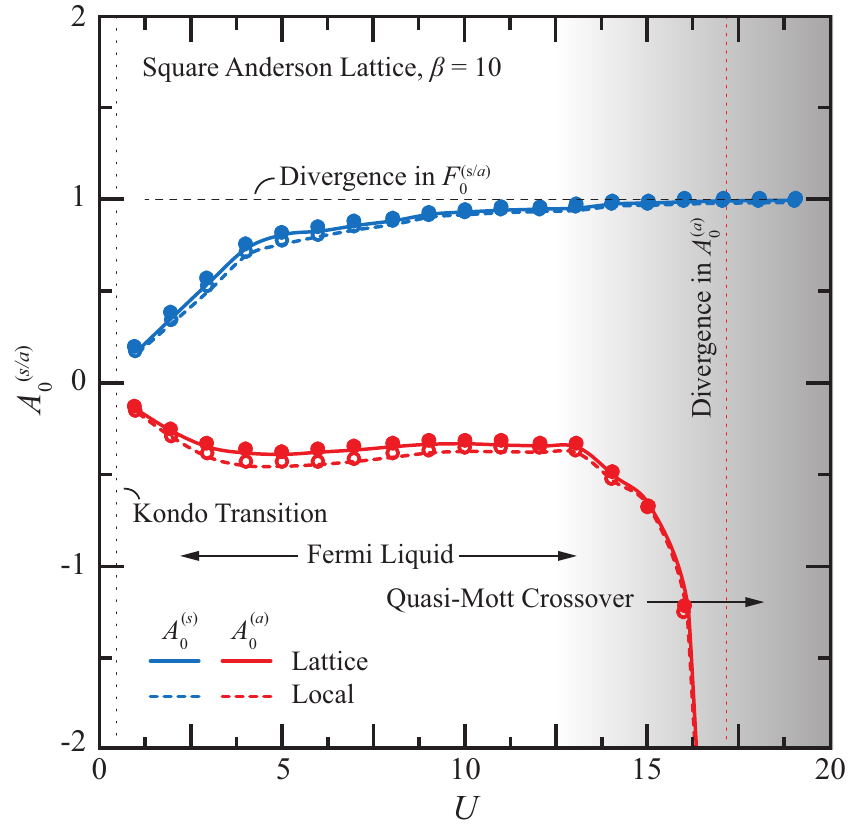}
\caption{\textcolor{\changescolor}{
Fermi liquid $A_0$ parameters of the local and lattice $f$ state in the square Anderson lattice as a function of $U$ at $\beta=10$.  As the lattice enters the quasi-Mott insulating phase, the antisymmetric parameters diverge (dotted red line) and the symmetric parameters converge to unity. This is qualitatively identical to the behavior in Hubbard model during a Mott crossover. We indicate that behavior in this figure with transition from a white to gray background. The $A_0=1$ line across which the $F_0$ parameters diverge is shown for reference. 
}
}
\label{fig:fl parameters A}
\end{figure}

\begin{figure}
\includegraphics[scale=1]{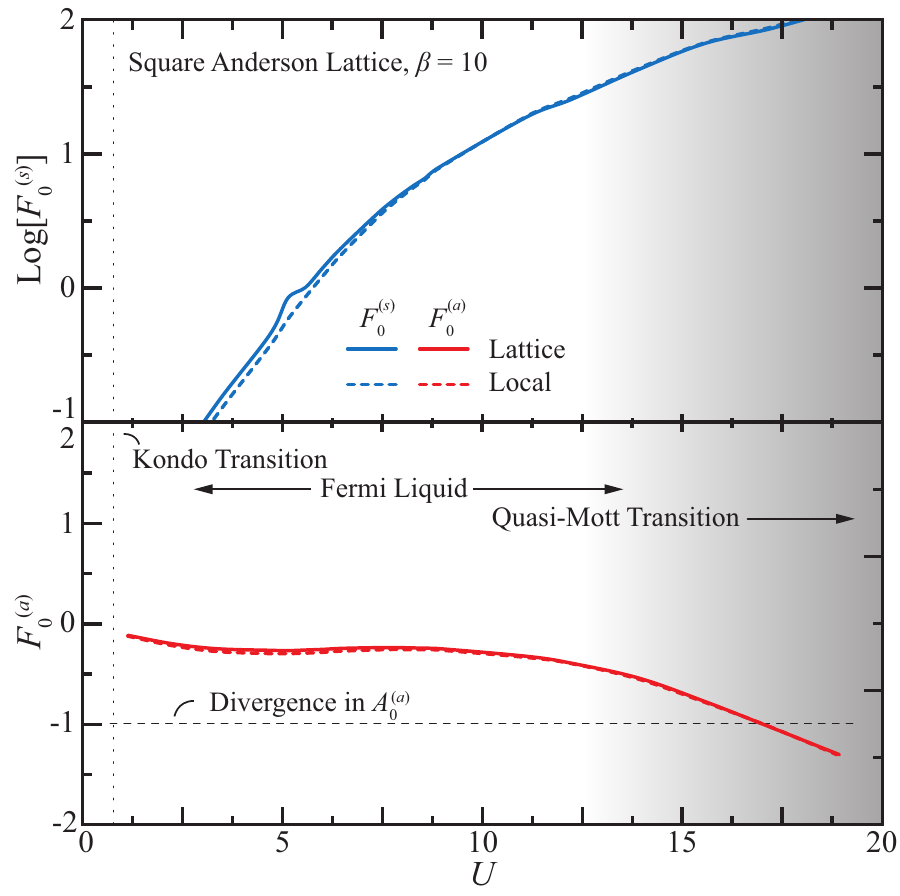}
\caption{
\textcolor{\changescolor}{
Fermi liquid $F_0$ parameters of the local and lattice $f$ state in the square Anderson lattice as a function of $U$ at $\beta=10$. The symmetric parameter goes towards infinity during the quasi-Mott transition, as in the Hubbard model during a Mott transition or crossover. The growth is relatively slow in comparison and $A_0^{(s)}$ remains sufficiently far from unity for us to avoid issues with the error.  The antisymmetric parameter also goes towards $-\infty$ during the quasi-Mott transition, but its growth is markedly slower than in the Hubbard model and there is no indication that a true divergence will occur, as expected. The $F_0=-1$ line across which $A_0$ diverges is shown for reference. We also indicate the quasi-Mott transition in this figure with a transition from a white to gray background.
}
}
\label{fig:fl parameters F}
\end{figure}

Let us examine the Fermi liquid parameters in the square Anderson lattice. Figures \ref{fig:fl parameters A} and \ref{fig:fl parameters F} show the Fermi liquid parameters in the square Anderson lattice for variations in the interaction strength. Interestingly, the $A_0^{(s/a)}$ of the Anderson lattice are qualitatively similar to those of the Hubbard lattice. That is, the symmetric parameters converge to unity and the asymmetric parameters diverge at large $U$. This is surprising, as the Anderson lattice does not undergo a true Mott transition. Instead, it becomes a quasi-Mott $f$-insulator with some spectral weight remaining between the two Mott bands. (See Fig. \ref{fig:phase diagram 1} for an example of the spectral functions in this regime.) Moreover,the square Anderson lattice does not undergo a magnetic transition, so this divergence does not have physical meaning. As in the Hubbard model, it must therefore indicate a breakdown of the theory.

\begin{figure}
\includegraphics[scale=1]{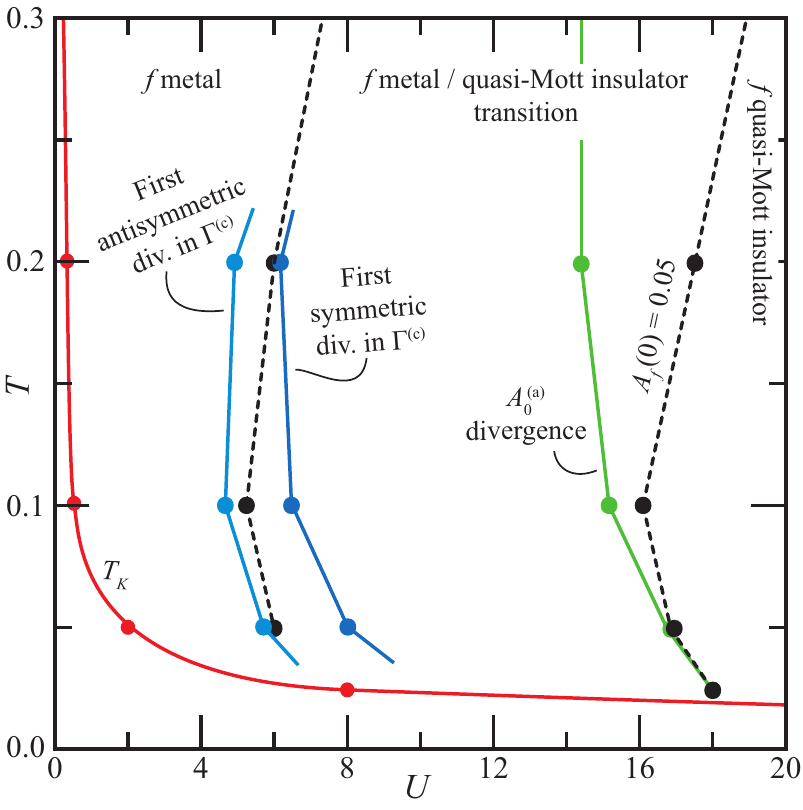}
\caption{The phase diagram of the square Anderson lattice, the first symmetric and antisymmetric divergence lines for the irreducible vertex, and the only divergence line of the Fermi liquid parameters [$A_0^{(a)}$]. The first irreducible vertex divergences occur near the start of the quasi-Mott transition in the $f$ states, while the Fermi liquid divergence occurs, at low temperature, when the $f$-states become nearly Mott insulating. We define the onset of the quasi-Mott insulating state by the location where the $f$-state spectral function $A_f(\omega=0)=0.05$.
}
\label{fig:fl parameters phase diagram}
\end{figure}

If we overlay the phase diagram of the square Anderson lattice with this divergence, as in Fig. \ref{fig:fl parameters phase diagram}, we see that at low temperatures the divergence occurs precisely when we enter the quasi-Mott insulator regime. Note, however, that there is not a clear boundary between the regimes: The spectral function never becomes truly Mott insulating, there is no divergence in the self-energy, and there are no signs that we have entered a Mott regime.  Instead, the quasiparticle peak is gradually suppressed, but never disappears.  In Fig. \ref{fig:fl parameters phase diagram}, we have somewhat arbitrarily selected $A_f(\omega=0)=0.05$ as the boundary between the transitioning and quasi-insulating phases, and this selection coincidentally aligns the ``phase boundary'' and the divergence in $A_0^{(a)}$. Still, we can connect this quasi-transition with the divergence. That is, the divergence is a sign that a substantial amount of Mott physics are present, that the Fermi-liquid picture is no longer complete, and that the Fermi-liquid theory has broken down. 

Finally, let us emphasize again that the behavior of the Fermi liquid is not correlated to the divergences of the irreducible vertex. From Fig. \ref{fig:fl parameters phase diagram}, we see that these divergences occur well before the divergence of $A_0^{(a)}$. Furthermore, they occur in the charge channel, whereas the divergence in the quasiparticle vertex occurs only in the spin channel (antisymmetric parameter).  Most importantly, the many divergences in the irreducible vertex in the charge channel do not prevent our perturbation theory based on this vertex from accurately capturing breakdown of Fermi-liquid theory as one enters a Mott or quasi-Mott regime. This is the most important result of this study. It shows that the divergences of the two-particle irreducible vertex do not indicate that there is a breakdown in the perturbation theory, just as divergences in the the self-energy do not indicate a breakdown of the perturbation theory. 

\subsection{Structure of the Fermi liquid vertex}
\label{sec:structure fl vertex}

Now let us examine the structure of the quasiparticle and incoherent vertices in the charge and spin channels. In particular, let us show why the charge channel objects cannot diverge unless the susceptibility diverges; and let us show why the spin channel objects can diverge even when the susceptibility does not. We begin with the charge channel. 

\subsubsection{Charge channel}

\begin{figure}
\includegraphics[scale=1]{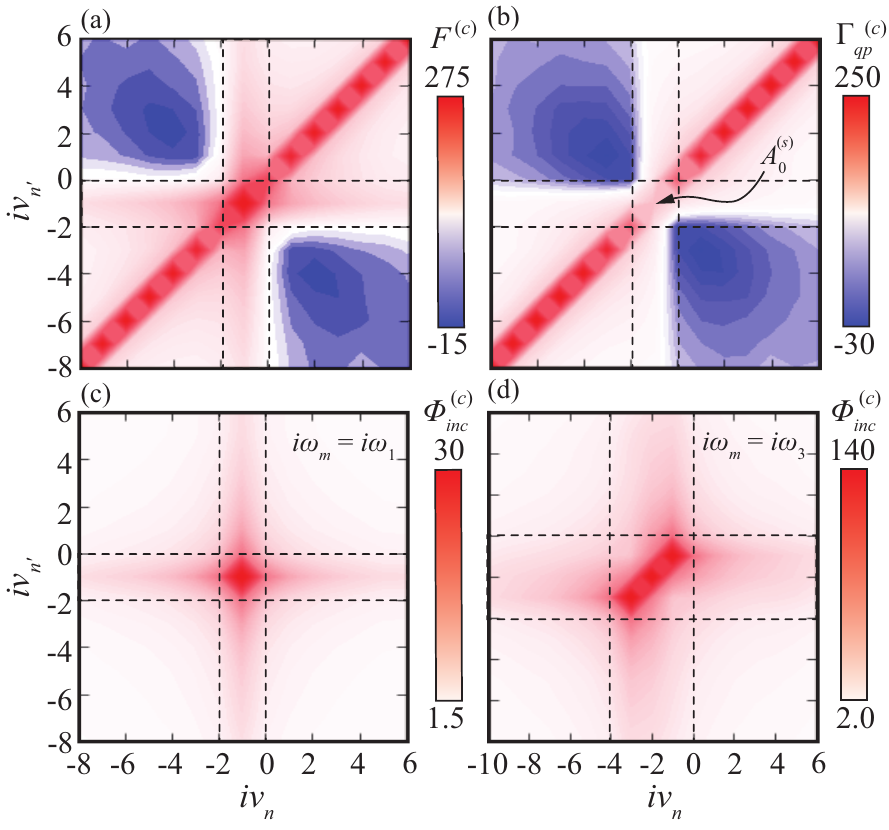}
\caption{(a) The full vertex, (b) quasiparticle vertex, and (c) incoherent vertex at $U=7$ and $\beta=10$  for the first non-zero bosonic frequency in the square Anderson lattice with $\bm{q}=\bm{0}$. The quasiparticle vertex contains the asymmetric decaying components and asymptotic diagonal ($\nu\simeq\nu'$) structure of the full vertex, whereas the incoherent vertex contains the symmetric decaying components and the asymptotic cross ($\nu=-\omega/2$ and $\nu'=-\omega/2$) and uniform background structures. The incoherent vertex also absorbs those elements of the diagonal structure which fall within the low-frequency regime defined by the coherent bubble ($-\omega<\nu,\nu'<0$, dashed lines), as shown more clearly in (d) which displays the incoherent vertex at the third non-zero bosonic frequency. (b) highlights the Fermi liquid parameter, which is proportional to the central element of $\Gamma_{qp}^{\omega_1}$.  }
\label{fig:qp vertex}
\end{figure}

Figure \ref{fig:qp vertex} shows a typical set of the full vertex, and its coherent and incoherent parts. Figure \ref{fig:qp vertex}(b) highlights that the central element of the quasiparticle vertex $\bm{\Gamma_}{qp}^{i\omega_1}$ is proportional to the Fermi liquid parameter $A_0^{(s)}$. [See Eq. \ref{eq:fermi liquid parameters}.] For the parameters chosen in this figure, the irreducible vertex has undergone multiple divergences; however, no divergence is experienced by the full, coherent, or incoherent vertices. Indeed, we find no divergence and no indication that such a divergence is possible unless the lattice susceptibility also diverges. Let us support this claim through a careful examinations of the structures which exist in the full, coherent, and incoherent vertices.

\begin{figure}
\includegraphics[scale=1]{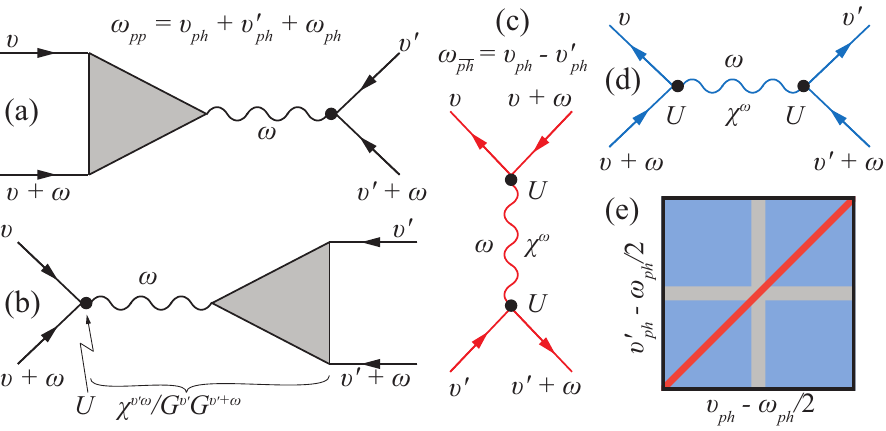}
\caption{(a-d) Diagrams responsible for the asymptotic structures of the full vertex, and (e) an illustration of the respective diagrams, color coded to match the diagrams. These diagrams and their colors are as follows: (a,b) The particle-particle kernel-2 diagrams ($pp$, gray). The (c) transverse particle-hole ($\overline{ph}$, red) and (d) particle-hole kernel-1 diagrams ($ph$, blue). Here $\chi^\omega$ is a single-time susceptibility and $\chi^{\omega\nu}$ is a two-time susceptibility. For a more thorough discussion, we refer the reader to Ref. [\citen{Kaufmann2017}].}
\label{fig:asymptotic diagrams}
\end{figure}

As one might expect from Eq. (\ref{eq:latt:full vertex to FL vertex}), the coherent vertex strongly resembles the full vertex as $\omega\rightarrow0$. However, there are major differences, particularly when one examines the asymptotic structure of the full vertex, i.e., those structures which do not decay as $\nu,\nu'\rightarrow\infty$.  
The full vertex shown in Fig. \ref{fig:qp vertex}(a) contains asymptotic structures on the main diagonal defined by $\nu=\nu'$ and on the cross defined by the four lines $\nu=\nu'=(\nu+\omega)=(\nu'+\omega)=0$ (the dashed lines in Fig. \ref{fig:qp vertex}). It also contains the constant background term. Of these structures, the incoherent vertex contains the background, the cross, and the portion of the diagonal that lies within the low-frequency domain, $\omega<\nu<0$; and the coherent vertex contains the remaining, high-frequency portion of the diagonal. These structures are generated by diagrams like those illustrated in Fig. \ref{fig:asymptotic diagrams}\cite{Wentzell2016,Kaufmann2017}. These structures can be linked to a bosonic exchange,\cite{Krien2019} 
 and no divergence can arise in these structures unless a local observable, e.g., $G_loc(i\nu)$ or $\bm{\chi}^\omega_{loc}$, diverges. Therefore, they cannot be responsible for a divergence in the coherent vertex unless there is a corresponding divergence in the impurity observables.

We classify the decaying structures according to their symmetry. The incoherent vertex contains the symmetric decaying structures, and the coherent vertex contains the antisymmetric and asymmetric decaying structures of the full vertex. These observations hold for both the PAM and Hubbard model regardless of the location in the phase diagram, bosonic frequency,  
$\omega$, or wavevector, $\bm{q}$.  Again, the full vertex cannot diverge unless the lattice susceptibility also diverges. Therefore, if the coherent vertex diverges and the full vertex does not, then the incoherent vertex must exhibit a corresponding divergence. However, the symmetry of the decaying structures are different in the coherent and incoherent vertices, preventing this balance. Therefore, we can conclude that the coherent vertex does not diverge unless the lattice susceptibility, a physical observable, also diverges. 

Finally, we note that these observations hold in the Hubbard lattice, see Appendix \ref{app:supp results} for a depiction of the coherent and incoherent parts of the full vertex in the Hubbard lattice. 

\subsubsection{Spin Channel}

Having now examined why the symmetric Fermi liquid parameter does not diverge, let us examine why the antisymmetric Fermi liquid parameter does diverge even as the spin susceptibility does not. To answer this question, we focus on the square Hubbard lattice, which exhibits divergences in the antisymmetric Fermi liquid parameter (and thus the Fermi liquid vertex in the spin channel) but not the spin susceptibility. 

\begin{figure}
\includegraphics[scale=1]{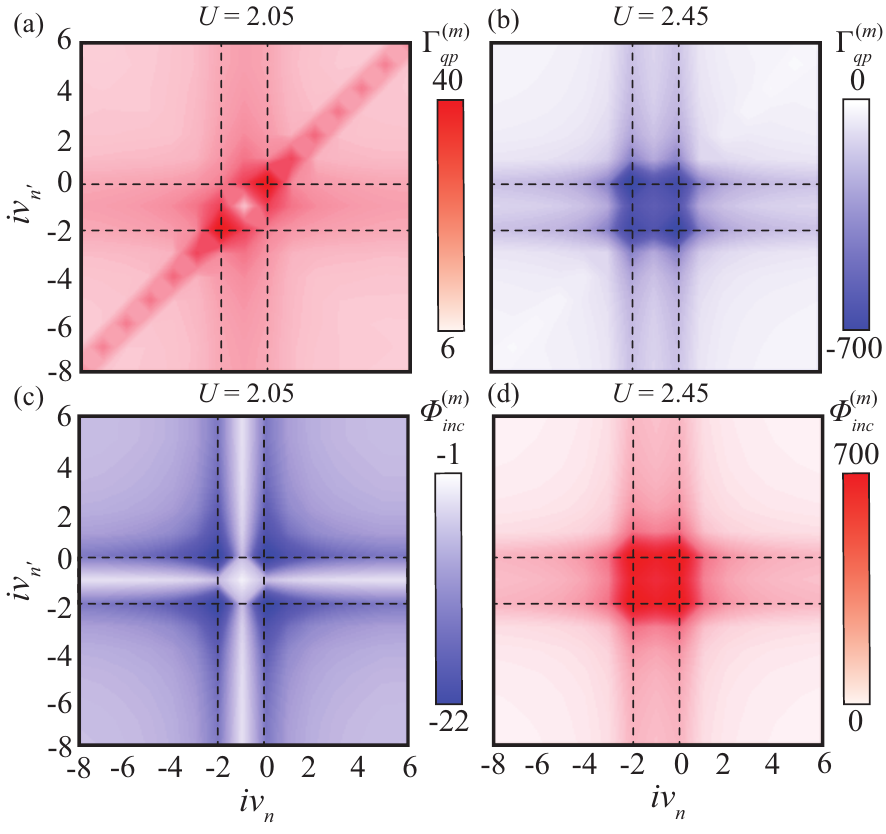}
\caption{Quasiparticle (a,b) and incoherent (c,d) vertices before (a,c) and after (b,d) the Mott MIT in the Square Hubbard lattice at $\beta=10$. The full vertex is dominated by the coherent part before the divergence at $U=2.4$, with only a small part of the constant background diagrams contained in the incoherent vertex. Near and after the divergence, the cross structure dominates in both coherent and incoherent vertices. These structures have opposite sign and cancel out, such that the full vertex does not diverge. }
\label{fig:fl vertex spin}
\end{figure}

The quasiparticle and incoherent vertices are shown in Fig. \ref{fig:fl vertex spin} on either side of the divergence (and MIT). As shown, the asymptotic structures of the full vertex are not segregated into coherent and incoherent structures in the spin channel like they are in the charge channel. Indeed, the coherent (quasiparticle) vertex and full vertex are nearly identical at low $U$, and the coherent vertex contains what appears to be all symmetric and asymetric components of the full vertex. In contrast, the incoherent vertex primarily contains some portion of the cross and constant background structures. 

While this incoherent vertex does suppress the cross and background structures, the  symmetry arguments used to explain why the Fermi liquid vertex does not diverge in the charge channel do not hold in the spin channel. For example, a divergence in the cross structure of the quasiparticle vertex can be compensated by a corresponding and opposing divergence in the cross structure of the incoherent vertex. This is precisely what happens in the square Hubbard lattice. That is, close to the MIT (and divergence), the cross structures of both quasiparticle and incoherent vertices grow very large with opposite sign and approximately equal magnitude, as shown in Fig. \ref{fig:fl vertex spin}. The comparatively small difference between the two cross structures leaves behind the cross structure of the full vertex, which does not diverge. 

Qualitatively similar results are found in the Anderson lattice, as shown in Appendix \ref{app:supp results}. 

It is not clear at this point why the structures in the spin and channel are not separated and why the structures in charge channel are. This is left for future study. For now, let us discuss how the Fermi liquid vertex can be used to investigate the coherent and incoherent parts of the susceptibility. 

\subsection{Coherent and incoherent susceptibilities}

One can use the quasiparticle vertex to examine the coherent part of the vertex-connected susceptibility, $\bm{\chi}^{q}_{qp}$, where $ \bm{\chi}^{q}_{qp} =  \bm{\chi}^{q}_0  \bm{\Gamma}^{q}_{qp}   \bm{\chi}^{q}_0$. Let us examine the coherent and incoherent parts of the lattice charge susceptibility in both the square Anderson and Hubbard lattices. We begin with the square Anderson lattice. 

\subsubsection{PAM}

\begin{figure}
\includegraphics[scale=1]{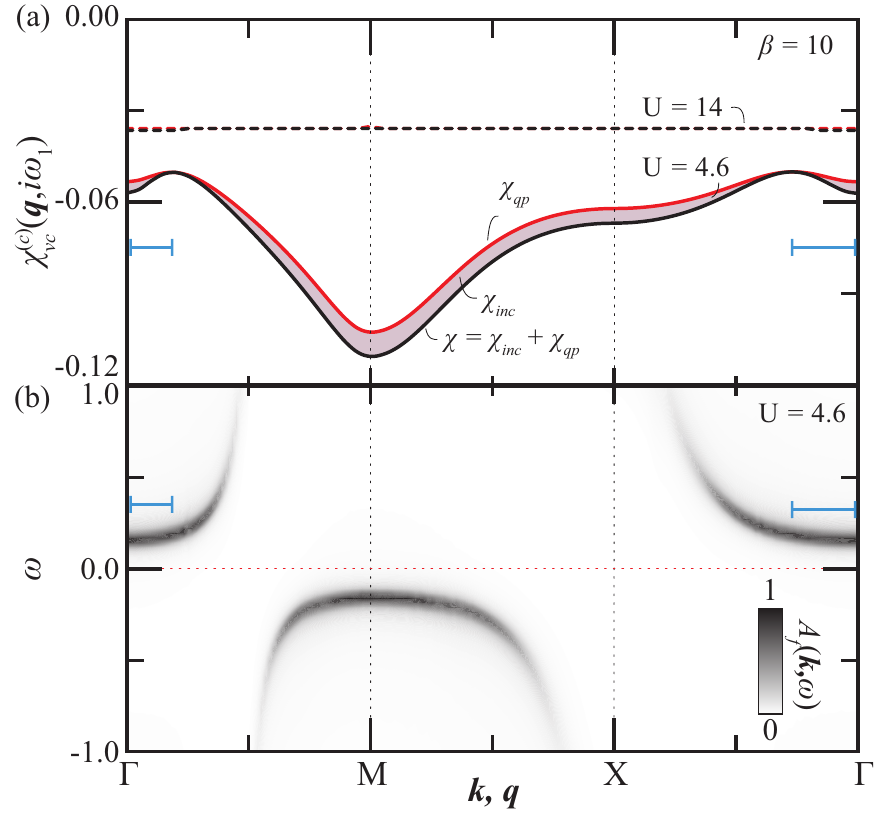}
\caption{(a) The  vertex-connected part of the lattice susceptibility and its coherent part along the high-symmetry lines of the BZ at the first bosonic frequency. The incoherent part is shown as the difference between these curves. (b) The spectral function of the $f$ states along the same high-symmetry lines. The purely coherent  dip in the susceptibility is connected to the intraband excitations, whereas the peak at X is connected to the interband excitations.  }
\label{fig:coherent susceptibility}
\end{figure}

Figure \ref{fig:coherent susceptibility}(a) shows the decomposition of the  vertex-connected part of the susceptibility into coherent and incoherent parts in the transitioning phase. (Figure \ref{fig:coherent susceptibility}(b) shows distinct valence and conduction quasi-particle bands in the $f$ states. However, substantial spectral weight remains at the Fermi level.) As shown, the coherent part of the susceptibility dominates the incoherent part. Moreover, both coherent and incoherent parts have the same shape: a (negative) peak at M, a plateau at X, and a dip in a ring around $\Gamma$. Interestingly, the incoherent part completely vanishes on this ring around the $\Gamma$-point, such that the susceptibility is completely coherent. Far from this ring, however, the coherent part contributes a nearly constant  fraction to the total susceptibility. This behavior is replicated in the Hubbard model in the metallic phase, as we will discuss in Sec. \ref{sec:hubbard qp susc}. 

\textcolor{\changescolor}{
Let us take a moment to note that the total susceptibility (the connected and bubble parts) vanishes as $\bm{q}\rightarrow\bm{0}$ due to the conservation of charge. However, the connected and bubble parts themselves do not. Here and throughout this study we focus only on the  vertex-connected part of the susceptibility, as this is the part which depends upon the vertex functions which are diverging and in which we are interested. }

Now let us examine the ring around the zone-center wherein the incoherent part vanishes. While the incoherent vertex is suppressed on this ring, it does not vanish. Instead, the incoherent part of the susceptibility vanishes because the cross structure (which suppresses the susceptibility) and the diagonal and symmetric decaying structures (which enhance the susceptibility) balance with each other on this ring. Inside the ring, the diagonal structure dominates; outside the ring, the symmetric decaying structures dominate.  The dip in the coherent susceptibility arises through a similar mechanism. However, the asymmetric decaying structures (i.e., those structures which exist within the region of the cross structure) do not dominate the coherent vertex. Therefore, they cannot fully suppress the contributions from the diagonal structure, which does dominate the vertex.

More physically, we can connect this ring of suppressed susceptibility to the intraband interactions: Figure \ref{fig:coherent susceptibility} shows that this ring corresponds to the trough of the conduction and valence bands. 

\subsubsection{Hubbard}
\label{sec:hubbard qp susc}

\begin{figure}
 \includegraphics{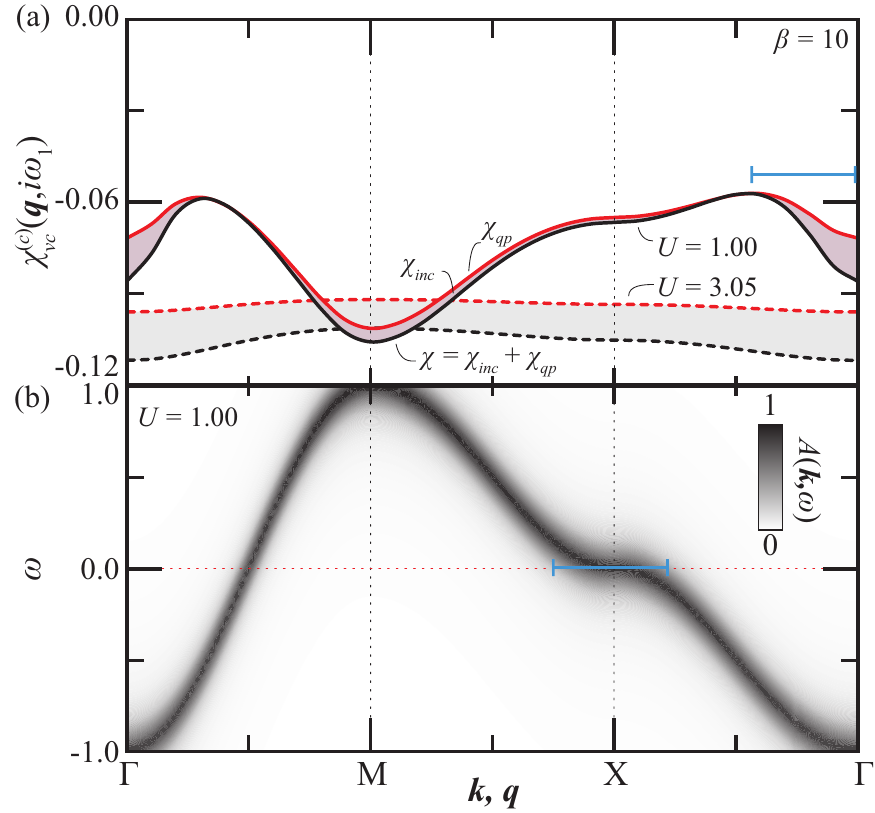}
 \caption{(a) The  vertex-connected part of the lattice charge susceptibility and its coherent part along the high-symmetry lines in the BZ for the first bosonic frequency.  (b) The spectral function along the same lines. In the metallic phase (solid lines), the susceptibility becomes both suppressed and purely coherent on a ring around the $\Gamma$-point. This coherent structure is connected to the nearly flat quasi-particle band crossing the Fermi-level at X. (The peak at M is connected to the X $\rightarrow$ X interactions.)  }
 \label{fig: coherent susceptibility hubbard}
 \end{figure}

Figure \ref{fig: coherent susceptibility hubbard}(a) shows the decomposition of the lattice charge susceptibility into coherent and incoherent parts for the metallic phase. The behavior is very similar to that of the PAM shown in Fig. \ref{fig:coherent susceptibility}. That is, the lattice susceptibility is entirely coherent on a ring around the $\Gamma$-point, where the coherent susceptibility is suppressed, and exhibits a peak at M and a plateau at X. Again, this purely coherent feature is connected with the features of the spectral function near the Fermi-surface, as shown in Fig. \ref{fig: coherent susceptibility hubbard}(b). That is, the phase space available for an ``intravalley'' interactions on the nearly flat quasi-particle crossing at X. In comparison, the peak at M is connected to the ``intervalley'' interactions: X$\rightarrow$X. 

Unsurprisingly, the Hubbard model in the Mott insulating phase is notably different from the PAM model in the quasi-Mott insulating phase: While both are relatively insensitive to the wavevector, the Hubbard model retains a substantial incoherent susceptibility in this phase, whereas the PAM model becomes almost completely coherent.  Additionally, the Hubbard model does exhibit some $\bm{q}$ dependence, with nearly the inverse structure to that seen in the metallic phase: a large dip at M, a plateau at X, and a peak at $\Gamma$. Of course, the Fermi liquid has vanished in the Hubbard model after the Mott transition, so the validity of the Fermi liquid theory is dubious at this point.

Note that we are only looking at the charge susceptibilities of the lattice. If we examine the spin susceptibilities, we see that coherent and incoherent parts of the susceptibility diverge near the Mott transition. This divergence begins at the M point before quickly spreading throughout the BZ. Of course, Landau Fermi liquid theory is primarily concerned with the low-energy ($\omega\rightarrow0$) and low-momentum transfer ($\bm{q}\rightarrow0$) behavior. Therefore, we must be cautious when investigating the response of the Fermi-liquid far from the zone-center. Let us return to more solid ground and summarize our major results.

\section{Conclusions}
\label{sec:conclusions}

In this study, we have examined the divergences of the irreducible vertex function and the two-particle response of the Fermi liquid in both the Anderson and Hubbard lattices. We have shown that these diverging vertex functions can be used to accurately capture the behavior of the Fermi liquid, indicating that the two-particle perturbation theories based around the irreducible vertex remain intact. 

In order to accomplish this, we build a quasiparticle vertex from the irreducible vertex by expanding it in ladders of the incoherent bubble. (Equivalently, the full vertex can be built via an expansion of the quasiparticle vertex in ladders of the coherent bubble.) Then, we show that the antisymmetric Fermi liquid $A_0$ parameters extracted from this vertex only diverge as one transitions from a Fermi-liquid to Mott physics regime, i.e., as the Fermi-liquid theory breaks down. Furthermore, we show that in the Fermi-liquid regime, these parameters behave as expected, regardless of the divergence of the underlying irreducible vertex. 

These results support that the symmetric divergences of the irreducible vertex are the physical result of a suppression of charge fluctuations, and not the non-physical consequence of the perturbation theory breaking down. Further support comes from our observation that the first symmetric divergences of the irreducible vertex occur near the maximum magnitude of the (negative) vertex-connected part of the charge susceptibility, which suppresses charge fluctuations. 

We also discuss the structure of the quasiparticle vertex and demonstrate that the structures of the full vertex in the charge channel can be cleanly separated into coherent and incoherent parts. However, we show that this is not true in the spin channel. This allows the associated Fermi liquid parameter, $A_0^{(a)}$, to diverge near a Mott MIT. In contrast, the division of structures in the charge channel prevents the associated Fermi liquid parameter,  $A_0^{(s)}$, from diverging. 

Finally, we use the quasiparticle vertex in order to compute the coherent and incoherent susceptibilities throughout the BZ. We show that features of the susceptibility may become completely coherent along particular curves (or surfaces) in the BZ, and that this coherence can be connected to the one-particle spectral function.  We show that this behavior is strongly $U$ dependent and that it vanishes in the Mott insulating regime, wherein one should not apply the Fermi-liquid theory. 

\section{Acknowledgements}

Gabi Kotliar is grateful to G. Sangiovanni and A. Toschi for a useful discussion.  This work was supported by the US Department of Energy, Office of Basic Energy Sciences as part of the Computational Materials Science Program.

%

\begin{appendix}

\section{Supplemental Results}\label{app:supp results}

In this Appendix we present a few results which may be helpful reference for the interested reader, but are either not new or else not unique between the two lattice models discussed. In particular, we show the  coherent and incoherent parts of the full vertex in the charge channel for the square Hubbard lattice and spin channel for the square Anderson lattice. We also present the charge and spin susceptibility in the square Anderson lattice.

 \begin{figure}[hb]
 \includegraphics{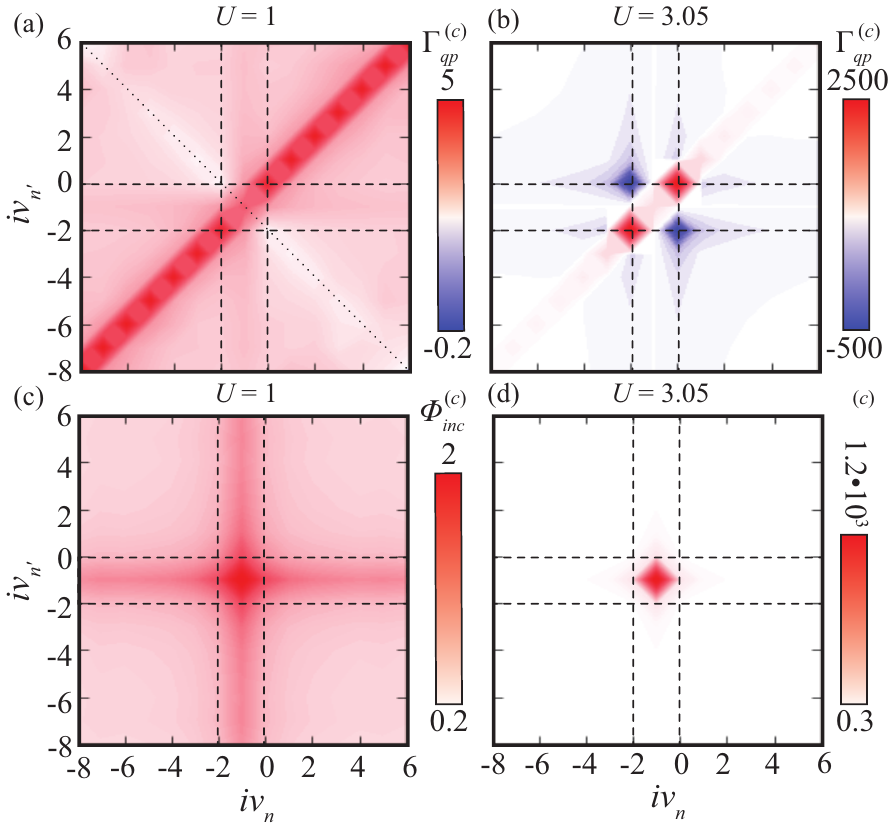}
 \caption{The coherent and incoherent parts of the full vertex in the metallic ($U=1$) and Mott insulating ($U=3.05$) phases of the square Hubbard lattice in the charge channel for the first bosonic frequency. Before the MIT, the typical asymptotic structures dominate the vertex: Diagonal, off-diagonal, cross, and constant background. After the MIT, the vertex is dominated by quickly decaying structures. }
 \label{fig: coherent vertex hubbard}
 \end{figure}

Figure \ref{fig: coherent vertex hubbard} presents the coherent and incoherent parts of the full vertex in the metallic and Mott insulating phases for the Hubbard Model in the charge channel. Contrast this with Fig. \ref{fig:qp vertex}, which shows similar results for the PAM. As in the PAM, the various asymptotic structures are divided between the coherent and incoherent parts; and the decaying structures appear to be separated according to their symmetry. For a thorough discussion of these structures, see Sec. \ref{sec:structure fl vertex}.

 \begin{figure}[hb]
 \includegraphics{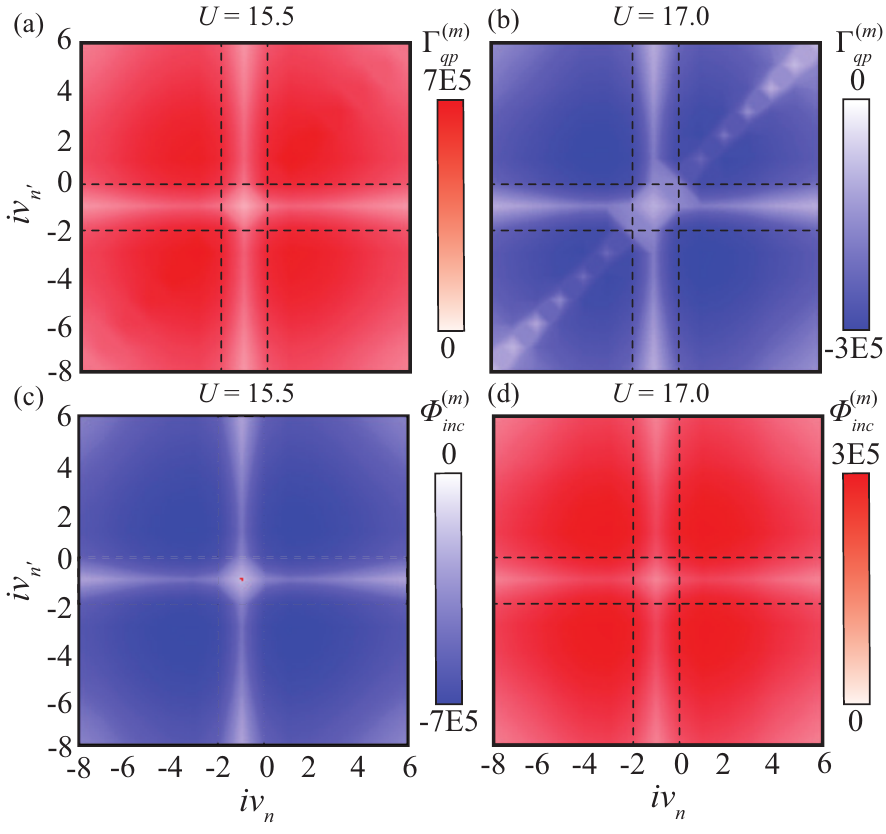}
 \caption{The coherent and incoherent parts of the full vertex in the spin channel near a divergence in the associated Fermi liquid parameter, $A_0^{(a)}$. The divergence is symmetric, and there is not the clear division of asymptotic structures seen in the charge channel.}
 \label{fig: coherent vertex spin}
 \end{figure}
 
 Figure \ref{fig: coherent vertex spin} presents the coherent and incoherent parts of the full vertex in the metallic and quasi-Mott insulating phases of the square Anderson lattice in the spin channel. Contrast this with Fig. \ref{fig:fl vertex spin}, which shows similar results for the Hubbard model.  As shown, and as in the Hubbard model, the divergence is created by a symmetric eigenvector, and there is not the clear division of asymptotic structures seen in the charge channel. For a more thorough discussion, see Sec. \ref{sec:structure fl vertex}.

 \begin{figure}[hb]
 \includegraphics{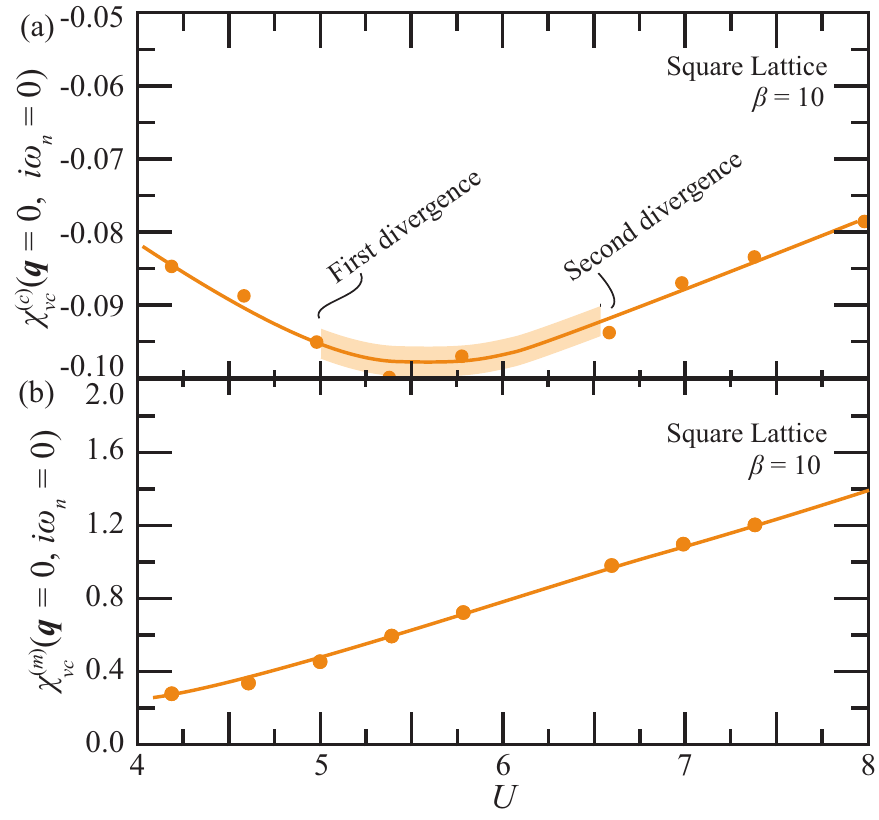}
 \caption{The (a) charge and (b) spin susceptibility of the $f$-state near the first divergence lines in the square Hubbard lattice at a few temperatures. The first symmetric divergence occurs near $\partial \chi^{(c)}_{vc} /\partial U \approx 0$, the maximum (negative) value of the vertex correction to the charge susceptibility. The first and second divergence divergence lines are indicated by the onset and termination of the shaded regions.  }
 \label{fig:susceptibility square PAM}
 \end{figure}
 
 Figure \ref{fig:susceptibility square PAM} shows the charge and spin susceptibilities in the square Anderson lattice. We present this figure in order to further test our claim that the first divergences occurs near $\partial \chi^{(c)}_{vc,f} /\partial U \approx 0$, c.f., Sec. \ref{sec:parquet results} and Fig. \ref{fig:susceptibility}. As shown, the divergences do occur near but not at the maximum magnitude of the negative $\chi^{(c)}_{vc,f}$. This helps to support the claim that the divergences are associated with the suppression of charge fluctuations.

\end{appendix}

\end{document}